\documentclass[preprint]{sigplanconf}
\makeatletter
\@ifundefined{lhs2tex.lhs2tex.sty.read}%
  {\@namedef{lhs2tex.lhs2tex.sty.read}{}%
   \newcommand\SkipToFmtEnd{}%
   \newcommand\EndFmtInput{}%
   \long\def\SkipToFmtEnd#1\EndFmtInput{}%
  }\SkipToFmtEnd

\newcommand\ReadOnlyOnce[1]{\@ifundefined{#1}{\@namedef{#1}{}}\SkipToFmtEnd}
\usepackage{amstext}
\usepackage{amssymb}
\usepackage{stmaryrd}
\DeclareFontFamily{OT1}{cmtex}{}
\DeclareFontShape{OT1}{cmtex}{m}{n}
  {<5><6><7><8>cmtex8
   <9>cmtex9
   <10><10.95><12><14.4><17.28><20.74><24.88>cmtex10}{}
\DeclareFontShape{OT1}{cmtex}{m}{it}
  {<-> ssub * cmtt/m/it}{}

\DeclareFontShape{OT1}{cmtt}{bx}{n}
  {<5><6><7><8>cmtt8
   <9>cmbtt9
   <10><10.95><12><14.4><17.28><20.74><24.88>cmbtt10}{}
\DeclareFontShape{OT1}{cmtex}{bx}{n}
  {<-> ssub * cmtt/bx/n}{}

\newcommand{\Conid}[1]{\mathit{#1}}
\newcommand{\Varid}[1]{\mathit{#1}}
\newcommand{\anonymous}{\kern0.06em \vbox{\hrule\@width.5em}}
\newcommand{\plus}{\mathbin{+\!\!\!+}}

\usepackage{polytable}

\@ifundefined{mathindent}%
  {\newdimen\mathindent\mathindent\leftmargini}%
  {}%

\def\resethooks{%
  \global\let\SaveRestoreHook\empty
  \global\let\ColumnHook\empty}
\newcommand*{\savecolumns}[1][default]%
  {\g@addto@macro\SaveRestoreHook{\savecolumns[#1]}}
\newcommand*{\restorecolumns}[1][default]%
  {\g@addto@macro\SaveRestoreHook{\restorecolumns[#1]}}
\newcommand*{\aligncolumn}[2]%
  {\g@addto@macro\ColumnHook{\column{#1}{#2}}}

\resethooks

\newcommand{\onelinecommentchars}{\quad-{}- }
\newcommand{\commentbeginchars}{\enskip\{-}
\newcommand{\commentendchars}{-\}\enskip}

\newcommand{\visiblecomments}{%
  \let\onelinecomment=\onelinecommentchars
  \let\commentbegin=\commentbeginchars
  \let\commentend=\commentendchars}

\newcommand{\invisiblecomments}{%
  \let\onelinecomment=\empty
  \let\commentbegin=\empty
  \let\commentend=\empty}

\visiblecomments

\newlength{\blanklineskip}
\newcommand{\hsindent}[1]{\quad}
\let\hspre\empty
\let\hspost\empty

\EndFmtInput
\makeatother
\ReadOnlyOnce{polycode.fmt}%
\makeatletter

\newcommand{\hsnewpar}[1]%
  {{\parskip=0pt\parindent=0pt\par\vskip #1\noindent}}

\newcommand{\hscodestyle}{}

\newcommand{\sethscode}[1]%
  {\expandafter\let\expandafter\hscode\csname #1\endcsname
   \expandafter\let\expandafter\endhscode\csname end#1\endcsname}

  {\par\noindent
   \advance\leftskip\mathindent
   \hscodestyle
   \let\\=\@normalcr
   \let\hspre\(\let\hspost\)%
   \pboxed}%
  {\endpboxed\)%
   \par\noindent
   \ignorespacesafterend}

  {\hsnewpar\abovedisplayskip
   \advance\leftskip\mathindent
   \hscodestyle
   \let\hspre\(\let\hspost\)%
   \pboxed}%
  {\endpboxed%
   \hsnewpar\belowdisplayskip
   \ignorespacesafterend}

  {\hsnewpar\abovedisplayskip
   \advance\leftskip\mathindent
   \hscodestyle
   \let\\=\@normalcr
   \(\pboxed}%
  {\endpboxed\)%
   \hsnewpar\belowdisplayskip
   \ignorespacesafterend}

\newcommand{\plainhs}{\sethscode{plainhscode}}

\plainhs

  {\hsnewpar\abovedisplayskip
   \advance\leftskip\mathindent
   \hscodestyle
   \let\\=\@normalcr
   \(\parray}%
  {\endparray\)%
   \hsnewpar\belowdisplayskip
   \ignorespacesafterend}

  {\parray}{\endparray}

  {\(\parray}{\endparray\)}

\def\codeframewidth{\arrayrulewidth}
\RequirePackage{calc}

  {\parskip=\abovedisplayskip\par\noindent
   \hscodestyle
   \arrayrulewidth=\codeframewidth
   \tabular{@{}|p{\linewidth-2\arraycolsep-2\arrayrulewidth-2pt}|@{}}%
   \hline\framedhslinecorrect\\{-1.5ex}%
   \let\endoflinesave=\\
   \let\\=\@normalcr
   \(\pboxed}%
  {\endpboxed\)%
   \framedhslinecorrect\endoflinesave{.5ex}\hline
   \endtabular
   \parskip=\belowdisplayskip\par\noindent
   \ignorespacesafterend}

\newcommand{\framedhslinecorrect}[2]%
  {#1[#2]}

  {\(\def\column##1##2{}%
   \let\>\undefined\let\<\undefined\let\\\undefined
   \newcommand\>[1][]{}\newcommand\<[1][]{}\newcommand\\[1][]{}%
   \def\fromto##1##2##3{##3}%
   }{\) }%

  {\let\orighscode=\hscode
   \let\origendhscode=\endhscode
   \def\endhscode{\def\hscode{\endgroup\def\@currenvir{hscode}\\}\begingroup}
   \orighscode\def\hscode{\endgroup\def\@currenvir{hscode}}}%
  {\origendhscode
   \global\let\hscode=\orighscode
   \global\let\endhscode=\origendhscode}%

\makeatother
\EndFmtInput
\ReadOnlyOnce{agda.fmt}%

\RequirePackage[T1]{fontenc}
\RequirePackage[utf8x]{inputenc}
\RequirePackage{ucs}
\RequirePackage{amsfonts}

\DeclareUnicodeCharacter{737}{\textsuperscript{l}}
\DeclareUnicodeCharacter{8718}{\ensuremath{\blacksquare}}
\DeclareUnicodeCharacter{8759}{::}
\DeclareUnicodeCharacter{9669}{\ensuremath{\triangleleft}}
\DeclareUnicodeCharacter{8799}{\ensuremath{\stackrel{\scriptscriptstyle ?}{=}}}
\DeclareUnicodeCharacter{10214}{\ensuremath{\llbracket}}
\DeclareUnicodeCharacter{10215}{\ensuremath{\rrbracket}}

\providecommand\textmu{$\mu$}

\renewcommand\Varid[1]{\mathord{\textsf{#1}}}
\let\Conid\Varid
\newcommand\Keyword[1]{\textsf{\textbf{#1}}}
\EndFmtInput

\usepackage{amsmath}
\usepackage{proof}
\usepackage{stmaryrd}
\usepackage{graphicx}
\usepackage{url}
\usepackage{color}
\usepackage[usenames,dvipsnames,svgnames,table]{xcolor}
\usepackage{ucs}
\usepackage[utf8x]{inputenc}
\usepackage{csquotes}

\DeclareUnicodeCharacter{7523}{\ensuremath{_r}}
\DeclareUnicodeCharacter{8343}{\ensuremath{_l}}
\DeclareUnicodeCharacter{8345}{\ensuremath{_n}}
\DeclareUnicodeCharacter{8337}{\ensuremath{_e}}
\DeclareUnicodeCharacter{8347}{\ensuremath{_s}}
\DeclareUnicodeCharacter{8348}{\ensuremath{_t}}
\DeclareUnicodeCharacter{7522}{\ensuremath{_i}}
\DeclareUnicodeCharacter{7525}{\ensuremath{_v}}
\DeclareUnicodeCharacter{10631}{\ensuremath{\llparenthesis}}
\DeclareUnicodeCharacter{10632}{\ensuremath{\rrparenthesis}}
\DeclareUnicodeCharacter{8599}{\ensuremath{\nearrow}}
\DeclareUnicodeCharacter{8598}{\ensuremath{\nwarrow}}
\DeclareUnicodeCharacter{8614}{\ensuremath{\mapsto}}
\DeclareUnicodeCharacter{8657}{\ensuremath{\Uparrow}}

\DeclareTextCommandDefault\textpi{\ensuremath{\pi}}
\DeclareTextCommandDefault\textlambda{\ensuremath{\lambda}}
\DeclareTextCommandDefault\textrho{\ensuremath{\rho}}
\DeclareTextCommandDefault\textGamma{\ensuremath{\Gamma}}
\DeclareTextCommandDefault\textiota{\ensuremath{\iota}}
\DeclareTextCommandDefault\textdelta{\ensuremath{\delta}}
\DeclareTextCommandDefault\textchi{\ensuremath{\chi}}
\DeclareTextCommandDefault\textXi{\ensuremath{\Xi}}
\DeclareTextCommandDefault\textxi{\ensuremath{\xi}}
\DeclareTextCommandDefault\textSigma{\ensuremath{\Sigma}}
\DeclareTextCommandDefault\textmu{\ensuremath{\mu}}
\DeclareTextCommandDefault\textalpha{\ensuremath{\alpha}}
\DeclareTextCommandDefault\textbeta{\ensuremath{\beta}}
\DeclareTextCommandDefault\textgamma{\ensuremath{\gamma}}
\DeclareTextCommandDefault\texteta{\ensuremath{\eta}}

\begin{document}

\setlength{\pdfpageheight}{\paperheight}
\setlength{\pdfpagewidth}{\paperwidth}

\conferenceinfo{ICFP '16}{September 18--24, 2016, Nara, Japan}
\copyrightyear{2016}
\copyrightdata{978-1-nnnn-nnnn-n/yy/mm} 
\copyrightdoi{nnnnnnn.nnnnnnn}          

\publicationrights{licensed}

\title{Embedding by Normalisation}
\authorinfo{Shayan Najd}
           {LFCS,\\ The University of Edinburgh}
           {sh.najd@ed.ac.uk}
\authorinfo{Sam Lindley}
           {LFCS,\\ The University of Edinburgh}
           {sam.lindley@ed.ac.uk}
\authorinfo{Josef Svenningsson}
           {Functional Programming Group,
            Chalmers University of Technology}
           {josefs@chalmers.se}
\authorinfo{Philip Wadler}
           {LFCS,\\ The University of Edinburgh}
           {wadler@inf.ed.ac.uk}
\maketitle

\begin{abstract}
This paper presents the insight that practical embedding techniques,
commonly used for implementing Domain-Specific Languages, correspond
to theoretical Normalisation-By-Evaluation (NBE) techniques, commonly
used for deriving canonical form of terms with respect to an
equational theory.

NBE constitutes of four components: a syntactic domain, a semantic
domain, and a pair of translations between the two.  Embedding also
often constitutes of four components: an object language, a host
language, encoding of object terms in the host, and extraction of
object code from the host.

The correspondence is deep in that all four components in embedding
and NBE correspond to each other. Based on this correspondence, this
paper introduces Embedding-By-Normalisation (EBN) as a principled
approach to study and structure embedding.

The correspondence is useful in that solutions from NBE can be
borrowed to solve problems in embedding. In particular, based on NBE
techniques, such as Type-Directed Partial Evaluation, this paper
presents a solution to the problem of extracting object code from
embedded programs involving sum types, such as conditional
expressions, and primitives, such as literals and operations on them.

\end{abstract}


\section{Introduction}
\label{sec:introduction}
Less is more sometimes. Compared to General-Purpose Languages (GPLs),
Domain-Specific Languages (DSLs) are smaller and simpler.
Unlike GPLs, DSLs are designed ground up to describe programs used in
a specific domain. DSLs are a powerful engineering
tool: DSLs abstract over domain-specific concepts and operations by
providing a set of primitives in the language.

\emph{Embedding} a DSL in a host GPL is by now well established as a family
of techniques for simplifying its implementation.
Embedded DSLs can reuse some of the existing
machinery implemented for their host language; for a particular
Embedded DSL (EDSL), one does not need to implement all the required
machinery.  For instance, by using quotations
\citep{QDSL,mainland-quoted}, macros \citep{racket}, overloading and
virtualisation of constructs \citep{ad-hoc,rompf2013scala}, or normal
techniques for modular programming \citep{1ML}, there is no
need for implementing a parser; by using higher-order abstract syntax
and piggybacking on module system of host , there is no need to
implement a name-resolver; and, by using mechanisms such as
Generalised Algebraic Data Types (GADTs) \citep{GADTs}, there is no
need to implement a type-checker.

Furthermore, as with any other host program, EDSL programs can often
integrate smoothly with other host programs, and reuse parts of the
ecosystem of the host language. Language-INtegrated Query (LINQ)
\citep{LINQ} is a well-known instance of such successful integration,
where SQL queries, as embedded DSL programs, are integrated with
programs in mainstream GPLs.

Although embedding can avoid remarkable implementation effort by
reusing the machinery available for the host language, it comes with
the expensive price of EDSLs losing their authentic identity.
Compared to stand-alone implementation, embedding is less flexible in
defining syntax and semantic of DSLs; more or less, syntax and
semantic of EDSLs often follow the ones of the host language.  There
are variety of smart and useful techniques to partially liberate EDSLs
from such restrictions (e.g., see \citet{QDSL, Definitional,
svenningsson:combiningJournal, Syntactic, scalalms}).
However, employing clever embedding techniques and
stacks of unconventional features available in the host language,
can lead to cryptic code sometimes. It becomes difficult
to distinguish an EDSL from the the host language, as the boundary
between an EDSL and its host language would not be entirely clear.
Implementation-based descriptions make EDSL rather
difficult to learn, not only for domain experts, whom traditionally
are assumed to be unfamiliar with the host language, but also for the
host language experts unfamiliar with the domain. Whole is nothing
without all its parts, and whole is greater than the sum its parts.

Unlike stand-alone languages that are often accompanied by a set of
formal descriptions, EDSLs are often presented by actual code in a
mainstream host language. Also, the embedding techniques themselves
are described in terms of a unique set of language features they
employ.  For instance in Haskell, a deep embedding means
datatypes in the host language are used for representing the syntax of EDSLs, and
functions (programs in general) over the datatypes are used for
defining semantics; or, in a final tagless embedding \citep{Tagless}
type-classes are used to define an interface representing syntax,
and instances of the type-classes are used for defining semantics.

Implementation-based descriptions make embedding techniques rather
difficult to learn as well: techniques vary greatly from one host
language to another, and even in a host language it is difficult to
compare techniques. As a result, existing techniques are hard to
scale. For instance, once one moves from embedding simpler DSLs to
DSLs with richer computational content, it becomes harder for
embedding to stay close to the intended syntax and semantic on one
hand, and reuse the host machinery on the other.
Would it not be convenient to have a more formal and
implementation-independent description of EDSLs and embedding
techniques? This paper is taking a few steps toward this goal.

The kind of principles and descriptions this paper is aiming for is
the ones of mathematical nature: abstract, insightful, and
simple. These are the kind of principles that have been guiding design
of functional programs since their dawn. One may argue these
principles are the ones that are discovered, as opposed to being
invented \cite{Wadler-2015}.

For instance, \citet[p. 513]{Tagless} observes that final
tagless embeddings are semantic algebras and form fold-like
structures. This observation has been explored further by \citet{Gibbons},
where they identify shallow embedding as algebras of folds over syntax
datatypes in deep embedding.  Decomposing embedding techniques into
well-know structures such as semantic algebras or folds is liberating:
embedding techniques can be studied independent of language features.
Semantic algebras and folds enjoy clear mathematical and formal
descriptions (e.g., via categorical semantics), hence establishing
correspondence between embedding and folds enables borrowing ideas
from other related fields.

In pursuit of a formal and implementation-independent description of
practical embedding techniques, this paper proposes
Embedding-By-Normalisation, EBN for short, as a principled approach to
study and structure embedding. EBN is based on a direct correspondence
between embedding techniques in practice and
Normalisation-By-Evaluation \citep{MartinLof,Berger} (NBE) techniques
in theory.

NBE is a well-studied approach (e.g., see
\citet{NBE-Cat,NBE-Sum,NBE-Untyped,Lindley05}) in proof theory and
programming semantics, commonly used for deriving canonical form of
terms with respect to an equational theory.  Decomposing embedding
techniques into the key structures in NBE is liberating: embedding
techniques can be studied independent of language features and
implementations. NBE enjoys clear mathematical and formal description,
and the connection between NBE and EDSLs allows for transferring
results from one field to the other, thereby strengthening both.  For
instance, in this paper we show how to use the NBE technique
Type-Directed Partial Evaluation (TDPE)\citep{TDPE} to extract object
code from host terms involving sums types, such as conditional
expressions, and primitives, such as literals and operations on
them. Although, there may exist various smart practical solutions to
the mentioned code extraction problem; at the time of writing this
paper, the process of code extraction for sum types and primitives is
considered an open theoretical problem in the EDSL community (see
\citet{Gill:CACM}).

The contributions of this paper are as follows:
\begin{itemize}
\item To characterise the correspondence between
      Normalisation-By-Evaluation (NBE) and embedding techniques
      (Section \ref{sec:NBE} and \ref{sec:EBN})
\item To introduce Embedding-By-Normalisation (EBN) as a principled
      approach to to study and structure embedding inspired by the
      correspondence to NBE (Section \ref{sec:NBE} and \ref{sec:EBN})
\item To propose a simple parametric model capturing a large and
      popular class of EDSLs, and introducing a series of EBN
      techniques for this model (Section \ref{sec:Type-Constrained})
\item To show how to extract code from embedded terms involving sum
      types, such as conditional expressions, as a by-product of EBN
      for above model involving sum types (Section \ref{sec:Sums})
\item To show how to extract code from embedded terms involving
      primitive values and operations, as a by-product of EBN
      for above model involving primitives (Section \ref{sec:Smart})
\item To show how EBN relates to some of the related existing
      techniques, and highlighting some insights when
      observing such techniques through EBN lens
      (Section \ref{sec:RelatedWork})
\end{itemize}

To stay formal and implementation-independent, the descriptions and
code in the main body of this paper is presented using type theory,
following a syntax similar to the one in the proof assistant Agda
\citep{Agda}.  Only a minimal set of language features is used, hoping
for the presentation to remain accessible to the readers familiar with
functional programming.  When inferrable from context, some
unnecessary implementation details, such as type instantiations or
overloading of constants, are intentionally left out of the code for
brevity.  The implementation concerns are addressed separately
throughout the paper.  Code and definitions presented in this paper
are implemented in Agda, and are available at
\url{https://github.com/shayan-najd/Embedding-By-Normalisation}.

\section{Normalisation-By-Evaluation}
\label{sec:NBE}
Normalisation-by-Evaluation (NBE) is the process of deriving canonical
form of terms with respect to an equational theory.The process of
deriving canonical forms is often referred to as normalisation, where
the canonical forms are refereed to as normal forms. NBE dates back to
\citet{MartinLof}, where he used a similar technique, although not by
its current name, for normalising terms in type theory.
\citet{Berger} introduced NBE as an efficient normalisation technique.
In the context of proof theory, they observed that the round trip of
first evaluating terms, and then applying an inverse of the evaluation
function, normalises the terms. Following \citet{Berger}, \citet{TDPE}
used NBE to implement an offline partial evaluator that only required
types of terms to partially evaluate them.

There are different approaches to normalisation.  One popular approach
to normalisation is reduction-based, where a set of rewrite rules are
applied exhaustively until they can no longer be applied.  In contrast
to reduction-based approaches, NBE is defined based on a pair of
well-known program transformations, instead of rewrite rules. For this
reason, NBE is categorised as a reduction-free normalisation process.

NBE constitutes of four components:
\begin{description}
\item [Syntactic Domain]
is the language of terms to be
normalised by a NBE algorithm.

\item [Semantic Domain]
is another language used in NBE, defining a model for
evaluating terms in the syntactic domain. Often the semantic domain
contains parts of the syntactic domain left uninterpreted. The
uninterpreted parts are referred to as the residual parts, and in their
presence the semantic model as the residualising model.

\item [Evaluation]
is the process of mapping terms in the syntactic domain to
the corresponding elements in the semantic domain. Despite the name,
the evaluation process in NBE is often quite different from the one in
the standard evaluators.  Although it is not necessarily required, the
evaluation process in NBE is often compositional.
In this paper, following the convention, evaluation functions are
denoted as \ensuremath{\Varid{⟦\char95 ⟧}}. In the typed variant of NBE, same notation is also
used to denote mapping of types in evaluation.

\item [Reification]
is the process of mapping (back) elements of semantic
domain to the corresponding terms in the syntactic domain.
In this paper, following the convention, reification functions are
denoted as \ensuremath{\Varid{↓}}.

\end{description}

More formally, an algorithm with NBE structure can be seen as an
instance of the following (dependent) record:
\begin{hscode}\SaveRestoreHook
\column{B}{@{}>{\hspre}l<{\hspost}@{}}%
\column{10}{@{}>{\hspre}l<{\hspost}@{}}%
\column{15}{@{}>{\hspre}l<{\hspost}@{}}%
\column{16}{@{}>{\hspre}l<{\hspost}@{}}%
\column{E}{@{}>{\hspre}l<{\hspost}@{}}%
\>[B]{}\Conid{NBE}\;\mathrel{=}\;\{\mskip1.5mu {}\<[10]%
\>[10]{}\Conid{Syn}\;{}\<[15]%
\>[15]{}\mathbin{:}\;\Conid{Type}\;\Varid{,}{}\<[E]%
\\
\>[10]{}\Conid{Sem}\;{}\<[15]%
\>[15]{}\mathbin{:}\;\Conid{Type}\;\Varid{,}{}\<[E]%
\\
\>[10]{}\Varid{⟦\char95 ⟧}\;{}\<[16]%
\>[16]{}\mathbin{:}\;\Conid{Syn}\;\Varid{→}\;\Conid{Sem}\;\Varid{,}{}\<[E]%
\\
\>[10]{}\Varid{↓}\;{}\<[15]%
\>[15]{}\mathbin{:}\;\Conid{Sem}\;\Varid{→}\;\Conid{Syn}\mskip1.5mu\}{}\<[E]%
\ColumnHook
\end{hscode}\resethooks

As mentioned, normalisation in NBE is the round trip of first
evaluating terms, and then reifying them back. Therefore,
normalisation in NBE is a mapping from syntactic domain
to syntactic domain:

\begin{hscode}\SaveRestoreHook
\column{B}{@{}>{\hspre}l<{\hspost}@{}}%
\column{E}{@{}>{\hspre}l<{\hspost}@{}}%
\>[B]{}\Varid{norm}\;\mathbin{:}\;\Conid{Syn}\;\Varid{→}\;\Conid{Syn}{}\<[E]%
\\
\>[B]{}\Varid{norm}\;\Conid{M}\;\mathrel{=}\;\Varid{↓}\;\Varid{⟦}\;\Conid{M}\;\Varid{⟧}{}\<[E]%
\ColumnHook
\end{hscode}\resethooks

In a typed setting, it is expected that transformations in NBE to
preserve types of the terms. More formally, an algorithm with NBE
structure in a typed setting can be seen as an instance of the
following (dependent) record, with the following normalisation
function:

\begin{hscode}\SaveRestoreHook
\column{B}{@{}>{\hspre}l<{\hspost}@{}}%
\column{15}{@{}>{\hspre}l<{\hspost}@{}}%
\column{20}{@{}>{\hspre}l<{\hspost}@{}}%
\column{21}{@{}>{\hspre}l<{\hspost}@{}}%
\column{E}{@{}>{\hspre}l<{\hspost}@{}}%
\>[B]{}\Conid{TypedNBE}\;\mathrel{=}\;\{\mskip1.5mu {}\<[15]%
\>[15]{}\Conid{Syn}\;{}\<[20]%
\>[20]{}\mathbin{:}\;\underline{\text{Type}}\;\Varid{→}\;\Conid{Type}\;\Varid{,}{}\<[E]%
\\
\>[15]{}\Conid{Sem}\;{}\<[20]%
\>[20]{}\mathbin{:}\;\underline{\text{Type}}\;\Varid{→}\;\Conid{Type}\;\Varid{,}{}\<[E]%
\\
\>[15]{}\Varid{⟦\char95 ⟧}\;{}\<[21]%
\>[21]{}\mathbin{:}\;\Varid{∀}\;\Conid{A.}\;\Conid{Syn}\;\Conid{A}\;\Varid{→}\;\Conid{Sem}\;\Conid{A}\;\Varid{,}{}\<[E]%
\\
\>[15]{}\Varid{↓}\;{}\<[20]%
\>[20]{}\mathbin{:}\;\Varid{∀}\;\Conid{A.}\;\Conid{Sem}\;\Conid{A}\;\Varid{→}\;\Conid{Syn}\;\Conid{A}\mskip1.5mu\}{}\<[E]%
\\[\blanklineskip]%
\>[B]{}\Varid{norm}\;\mathbin{:}\;\Varid{∀}\;\Conid{A.}\;\Conid{Syn}\;\Conid{A}\;\Varid{→}\;\Conid{Syn}\;\Conid{A}{}\<[E]%
\\
\>[B]{}\Varid{norm}\;\Conid{M}\;\mathrel{=}\;\Varid{↓}\;\Varid{⟦}\;\Conid{M}\;\Varid{⟧}{}\<[E]%
\ColumnHook
\end{hscode}\resethooks

Above, \ensuremath{\underline{\text{Type}}} denotes kind of object types. It is underlined to
contrast it with \ensuremath{\Conid{Type}} which is the kind of types in the metalanguage.

A valid NBE normalisation algorithm, both untyped and typed, should
guarantee that, (a) normalisation preserves the intended meaning of
the terms, and (b) normalisation derives canonical form of terms up to
certain congruence relation.


\subsection{A First Example}
\label{sec:CharsLists}
As the first example, consider the "hello world" of NBE, terms of the
following language:
\begin{hscode}\SaveRestoreHook
\column{B}{@{}>{\hspre}l<{\hspost}@{}}%
\column{E}{@{}>{\hspre}l<{\hspost}@{}}%
\>[B]{}\Varid{c}\;\Varid{∈}\;\Conid{Char}\;(\Varid{set}\;\Varid{of}\;\Varid{characters}){}\<[E]%
\\
\>[B]{}\Conid{L,M,N}\;\Varid{∈}\;\Conid{Chars}\;\Conid{::=}\;\Varid{ε₀}\;\mid \;\Conid{Chr}\;\Varid{c}\;\mid \;\Conid{M}\;\Varid{∙}\;\Conid{N}{}\<[E]%
\ColumnHook
\end{hscode}\resethooks

The language, referred to as Chars, consists of an empty string, a
string containing only one character, and concatenation of strings.

For example, the terms
\begin{hscode}\SaveRestoreHook
\column{B}{@{}>{\hspre}l<{\hspost}@{}}%
\column{E}{@{}>{\hspre}l<{\hspost}@{}}%
\>[B]{}\Conid{Chr}\;\text{\tt 'N'}\;\Varid{∙}\;(\Conid{Chr}\;\text{\tt 'B'}\;\Varid{∙}\;\Conid{Chr}\;\text{\tt 'E'}){}\<[E]%
\ColumnHook
\end{hscode}\resethooks
 and
\begin{hscode}\SaveRestoreHook
\column{B}{@{}>{\hspre}l<{\hspost}@{}}%
\column{E}{@{}>{\hspre}l<{\hspost}@{}}%
\>[B]{}(\Conid{Chr}\;\text{\tt 'N'}\;\Varid{∙}\;\Varid{ε₀})\;\Varid{∙}\;((\Conid{Chr}\;\text{\tt 'B'}\;\Varid{∙}\;\Varid{ε₀})\;\Varid{∙}\;(\Conid{Chr}\;\text{\tt 'E'}\;\Varid{∙}\;\Varid{ε₀})){}\<[E]%
\ColumnHook
\end{hscode}\resethooks
 both represent the string ``NBE".

The intended equational theory for this language is the one of
free monoids, i.e., congruence over the following equations:
\begin{center}
\begin{hscode}\SaveRestoreHook
\column{B}{@{}>{\hspre}l<{\hspost}@{}}%
\column{6}{@{}>{\hspre}l<{\hspost}@{}}%
\column{7}{@{}>{\hspre}l<{\hspost}@{}}%
\column{9}{@{}>{\hspre}l<{\hspost}@{}}%
\column{15}{@{}>{\hspre}l<{\hspost}@{}}%
\column{18}{@{}>{\hspre}l<{\hspost}@{}}%
\column{E}{@{}>{\hspre}l<{\hspost}@{}}%
\>[7]{}\Varid{ε₀}\;\Varid{∙}\;\Conid{M}\;{}\<[15]%
\>[15]{}\mathrel{=}\;{}\<[18]%
\>[18]{}\Conid{M}{}\<[E]%
\\
\>[6]{}\Conid{M}\;{}\<[9]%
\>[9]{}\Varid{∙}\;\Varid{ε₀}\;{}\<[15]%
\>[15]{}\mathrel{=}\;{}\<[18]%
\>[18]{}\Conid{M}{}\<[E]%
\\
\>[B]{}(\Conid{L}\;\Varid{∙}\;\Conid{M})\;\Varid{∙}\;\Conid{N}\;{}\<[15]%
\>[15]{}\mathrel{=}\;{}\<[18]%
\>[18]{}\Conid{L}\;\Varid{∙}\;(\Conid{M}\;\Varid{∙}\;\Conid{N}){}\<[E]%
\ColumnHook
\end{hscode}\resethooks
\end{center}
NBE provides a normalisation process to derive a Canonical form for
the terms with respect to above equational theory. If two terms
represent the same string, they have an identical canonical form.
For instance, the two example terms above normalise to the following
term in canonical form:
\begin{hscode}\SaveRestoreHook
\column{B}{@{}>{\hspre}l<{\hspost}@{}}%
\column{E}{@{}>{\hspre}l<{\hspost}@{}}%
\>[B]{}\Conid{Chr}\;\text{\tt 'N'}\;\Varid{∙}\;(\Conid{Chr}\;\text{\tt 'B'}\;\Varid{∙}\;(\Conid{Chr}\;\text{\tt 'E'}\;\Varid{∙}\;\Varid{ε₀})){}\<[E]%
\ColumnHook
\end{hscode}\resethooks

For a specific syntactic domain, in this case the Chars language,
there are different ways to implement a NBE algorithm, as
there are different semantic domains to choose from.
For pedagogical purposes, two distinct NBE algorithms are presented
for the Chars language based on two distinct semantic domains:
(1) lists of characters, and
(2) functions over the syntactic domain itself.

\subsubsection{Lists as Semantic}
The syntactic domain given to be the Chars language, and semantic
domain chosen to be a list of characters, the next step
for defining a NBE algorithm is defining an evaluation function:

\begin{hscode}\SaveRestoreHook
\column{B}{@{}>{\hspre}l<{\hspost}@{}}%
\column{10}{@{}>{\hspre}l<{\hspost}@{}}%
\column{E}{@{}>{\hspre}l<{\hspost}@{}}%
\>[B]{}\Varid{⟦\char95 ⟧}\;\mathbin{:}\;\Conid{Chars}\;\Varid{→}\;\Conid{List}\;\Conid{Char}{}\<[E]%
\\[\blanklineskip]%
\>[B]{}\Varid{⟦}\;\Varid{ε₀}\;{}\<[10]%
\>[10]{}\Varid{⟧}\;\mathrel{=}\;[]{}\<[E]%
\\
\>[B]{}\Varid{⟦}\;\Conid{Chr}\;\Varid{n}\;{}\<[10]%
\>[10]{}\Varid{⟧}\;\mathrel{=}\;[\mskip1.5mu \;\Varid{n}\;\mskip1.5mu]{}\<[E]%
\\
\>[B]{}\Varid{⟦}\;\Conid{M}\;\Varid{∙}\;\Conid{N}\;{}\<[10]%
\>[10]{}\Varid{⟧}\;\mathrel{=}\;\Varid{⟦}\;\Conid{M}\;\Varid{⟧}\;\plus \;\Varid{⟦}\;\Conid{N}\;\Varid{⟧}{}\<[E]%
\ColumnHook
\end{hscode}\resethooks

Evaluation defined above is a simple mapping from Chars terms to
lists, where empty string is mapped to empty list, singleton
string to singleton list, and concatenation of strings to
concatenation of lists.
For instance, the two example terms representing ``NBE" earlier are
evaluated to the list \ensuremath{\Varid{['N',}\;\text{\tt 'B'}\;\Varid{,}\;\text{\tt 'E'}\;\mskip1.5mu]}.

Above evaluation process is particularly interesting in that it is
compositional: semantic of a term is constructed from the semantic
of its subterms. Though compositionality is a highly desired property,
thanks to the elegant mathematical properties, the evaluation
process in NBE is not required to be compositional.
In fact, some evaluation functions cannot be defined compositionaly.
Compositionality of a function forces it to be expressible solely by
folds, and not every function can be defined solely in terms of folds.
For instance, evaluation that rely on some forms of global
transformations, sometimes cannot be expressed solely in terms of
folds.

The next step is to define a reification process:

\begin{hscode}\SaveRestoreHook
\column{B}{@{}>{\hspre}l<{\hspost}@{}}%
\column{13}{@{}>{\hspre}l<{\hspost}@{}}%
\column{E}{@{}>{\hspre}l<{\hspost}@{}}%
\>[B]{}\Varid{↓}\;\mathbin{:}\;\Conid{List}\;\Conid{Char}\;\Varid{→}\;\Conid{Chars}{}\<[E]%
\\[\blanklineskip]%
\>[B]{}\Varid{↓}\;[]\;{}\<[13]%
\>[13]{}\mathrel{=}\;\Varid{ε₀}{}\<[E]%
\\
\>[B]{}\Varid{↓}\;(\Varid{c}\;\Varid{∷}\;\Varid{cs})\;{}\<[13]%
\>[13]{}\mathrel{=}\;\Conid{Chr}\;\Varid{c}\;\Varid{∙}\;(\Varid{↓}\;\Varid{cs}){}\<[E]%
\ColumnHook
\end{hscode}\resethooks

Reification defined above is a simple mapping from lists to Chars
terms, where empty list is mapped to empty strings, cons of list head
to list tail to concatenation of the corresponding singleton string
to the reified string of tail.
For example, the list \ensuremath{\Varid{['N',}\;\text{\tt 'B'}\;\Varid{,}\;\text{\tt 'E'}\;\mskip1.5mu]} from earlier is reified to the
following term:
\begin{hscode}\SaveRestoreHook
\column{B}{@{}>{\hspre}l<{\hspost}@{}}%
\column{E}{@{}>{\hspre}l<{\hspost}@{}}%
\>[B]{}\Conid{Chr}\;\text{\tt 'N'}\;\Varid{∙}\;(\Conid{Chr}\;\text{\tt 'B'}\;\Varid{∙}\;(\Conid{Chr}\;\text{\tt 'E'}\;\Varid{∙}\;\Varid{ε₀})){}\<[E]%
\ColumnHook
\end{hscode}\resethooks
The reification function is also compositional.

Putting the pieces together normalisation function is defined as
usual:
\begin{hscode}\SaveRestoreHook
\column{B}{@{}>{\hspre}l<{\hspost}@{}}%
\column{E}{@{}>{\hspre}l<{\hspost}@{}}%
\>[B]{}\Varid{norm}\;\mathbin{:}\;\Conid{Chars}\;\Varid{→}\;\Conid{Chars}{}\<[E]%
\\
\>[B]{}\Varid{norm}\;\Conid{M}\;\mathrel{=}\;\Varid{↓}\;\Varid{⟦}\;\Conid{M}\;\Varid{⟧}{}\<[E]%
\ColumnHook
\end{hscode}\resethooks
As expected, above function derives canonical form of Chars terms.
For instance, we have
\begin{hscode}\SaveRestoreHook
\column{B}{@{}>{\hspre}l<{\hspost}@{}}%
\column{3}{@{}>{\hspre}l<{\hspost}@{}}%
\column{E}{@{}>{\hspre}l<{\hspost}@{}}%
\>[B]{}\Varid{norm}\;(\Conid{Chr}\;\text{\tt 'N'}\;\Varid{∙}\;(\Conid{Chr}\;\text{\tt 'B'}\;\Varid{∙}\;\Conid{Chr}\;\text{\tt 'E'}))\;{}\<[E]%
\\
\>[B]{}\hsindent{3}{}\<[3]%
\>[3]{}\mathrel{=}{}\<[E]%
\\
\>[B]{}\Varid{norm}\;((\Conid{Chr}\;\text{\tt 'N'}\;\Varid{∙}\;\Varid{ε₀})\;\Varid{∙}\;((\Conid{Chr}\;\text{\tt 'B'}\;\Varid{∙}\;\Varid{ε₀})\;\Varid{∙}\;(\Conid{Chr}\;\text{\tt 'E'}\;\Varid{∙}\;\Varid{ε₀})))\;{}\<[E]%
\\
\>[B]{}\hsindent{3}{}\<[3]%
\>[3]{}\mathrel{=}{}\<[E]%
\\
\>[B]{}\Conid{Chr}\;\text{\tt 'N'}\;\Varid{∙}\;(\Conid{Chr}\;\text{\tt 'B'}\;\Varid{∙}\;(\Conid{Chr}\;\text{\tt 'E'}\;\Varid{∙}\;\Varid{ε₀})){}\<[E]%
\ColumnHook
\end{hscode}\resethooks

\subsubsection{Functions as Semantic}
\label{sec:CharsHughes}
The syntactic domain given to be the Chars language, and semantic
domain now chosen to be functions over syntactic domain itself, the next step
for defining a NBE algorithm is defining an evaluation function:

\begin{hscode}\SaveRestoreHook
\column{B}{@{}>{\hspre}l<{\hspost}@{}}%
\column{10}{@{}>{\hspre}l<{\hspost}@{}}%
\column{E}{@{}>{\hspre}l<{\hspost}@{}}%
\>[B]{}\Varid{⟦\char95 ⟧}\;\mathbin{:}\;\Conid{Chars}\;\Varid{→}\;(\Conid{Chars}\;\Varid{→}\;\Conid{Chars}){}\<[E]%
\\[\blanklineskip]%
\>[B]{}\Varid{⟦}\;\Varid{ε₀}\;{}\<[10]%
\>[10]{}\Varid{⟧}\;\mathrel{=}\;\Varid{id}{}\<[E]%
\\
\>[B]{}\Varid{⟦}\;\Conid{Chr}\;\Varid{c}\;{}\<[10]%
\>[10]{}\Varid{⟧}\;\mathrel{=}\;\Varid{λ}\;\Conid{N}\;\Varid{→}\;(\Conid{Chr}\;\Varid{c})\;\Varid{∙}\;\Conid{N}{}\<[E]%
\\
\>[B]{}\Varid{⟦}\;\Conid{M}\;\Varid{∙}\;\Conid{N}\;{}\<[10]%
\>[10]{}\Varid{⟧}\;\mathrel{=}\;\Varid{⟦}\;\Conid{M}\;\Varid{⟧}\;\Varid{∘}\;\Varid{⟦}\;\Conid{N}\;\Varid{⟧}{}\<[E]%
\ColumnHook
\end{hscode}\resethooks

Evaluation defined above is a simple mapping from Chars terms to
functions from Chars to Chars, where empty string is mapped to
identity function, singleton string to a function that concatenates the
same singleton string to its input, and concatenation of strings to
function composition.
For instance, the two example terms representing ``NBE" earlier are
evaluated to the function
\begin{hscode}\SaveRestoreHook
\column{B}{@{}>{\hspre}l<{\hspost}@{}}%
\column{E}{@{}>{\hspre}l<{\hspost}@{}}%
\>[B]{}\Varid{λ}\;\Conid{N}\;\Varid{→}\;\Conid{Chr}\;\text{\tt 'N'}\;\Varid{∙}\;(\Conid{Chr}\;\text{\tt 'B'}\;\Varid{∙}\;(\Conid{Chr}\;\text{\tt 'E'}\;\Varid{∙}\;\Conid{N})){}\<[E]%
\ColumnHook
\end{hscode}\resethooks
Above evaluation is also compositional.

The next step is to define a reification process:

\begin{hscode}\SaveRestoreHook
\column{B}{@{}>{\hspre}l<{\hspost}@{}}%
\column{E}{@{}>{\hspre}l<{\hspost}@{}}%
\>[B]{}\Varid{↓}\;\mathbin{:}\;(\Conid{Chars}\;\Varid{→}\;\Conid{Chars})\;\Varid{→}\;\Conid{Chars}{}\<[E]%
\\[\blanklineskip]%
\>[B]{}\Varid{↓}\;\Varid{f}\;\mathrel{=}\;\Varid{f}\;\Varid{ε₀}{}\<[E]%
\ColumnHook
\end{hscode}\resethooks

Reification defined above is very simple: it applies semantic
function to empty string.
For example, the function
\begin{hscode}\SaveRestoreHook
\column{B}{@{}>{\hspre}l<{\hspost}@{}}%
\column{E}{@{}>{\hspre}l<{\hspost}@{}}%
\>[B]{}\Varid{λ}\;\Conid{N}\;\Varid{→}\;\Conid{Chr}\;\text{\tt 'N'}\;\Varid{∙}\;(\Conid{Chr}\;\text{\tt 'B'}\;\Varid{∙}\;(\Conid{Chr}\;\text{\tt 'E'}\;\Varid{∙}\;\Conid{N})){}\<[E]%
\ColumnHook
\end{hscode}\resethooks
from earlier is reified to the following term:
\begin{hscode}\SaveRestoreHook
\column{B}{@{}>{\hspre}l<{\hspost}@{}}%
\column{E}{@{}>{\hspre}l<{\hspost}@{}}%
\>[B]{}\Conid{Chr}\;\text{\tt 'N'}\;\Varid{∙}\;(\Conid{Chr}\;\text{\tt 'B'}\;\Varid{∙}\;(\Conid{Chr}\;\text{\tt 'E'}\;\Varid{∙}\;\Varid{ε₀})){}\<[E]%
\ColumnHook
\end{hscode}\resethooks
Reification is also obviously compositional.

Normalisation function is defined as usual.  However, the
normalisation process based on function semantics is more efficient
compared to the one based on list semantics.  Essentially, the
evaluation process using the semantic domain \ensuremath{\Conid{Char}\;\Varid{→}\;\Conid{Char}}, evaluates
terms using an efficient representation of lists, known as Hughes
lists \citep{hughes1986novel}.

\subsubsection{Observation}
\label{sec:ADT}
For this example, three domains are explicitly discussed: Chars
syntactic domain, semantic domain based on normal lists and semantic
domain based on Hughes lists.  There is also a fourth domain implicit
in the discussion: the syntactic domain of canonical forms, which is a
subset of the syntactic domain. For Chars language, terms in canonical
form are of the following grammar:
\begin{hscode}\SaveRestoreHook
\column{B}{@{}>{\hspre}l<{\hspost}@{}}%
\column{E}{@{}>{\hspre}l<{\hspost}@{}}%
\>[B]{}\Varid{c}\;\Varid{∈}\;\Conid{Char}\;(\Varid{set}\;\Varid{of}\;\Varid{characters}){}\<[E]%
\\
\>[B]{}\Conid{N}\;\Varid{∈}\;\Conid{CanonicalChars}\;\Conid{::=}\;\Varid{ε₀}\;\mid \;\Conid{Chr}\;\Varid{c}\;\Varid{∙}\;\Conid{N}{}\<[E]%
\ColumnHook
\end{hscode}\resethooks

For instance, the example canonical form derived earlier follows the
above grammar.

Compared to Chars, the grammar of canonical forms is less flexible but
more compact: it is easier to program in Chars, but it is also harder
to analyse programs in Chars. At the cost of implementing a
normalisation process, like the two NBE algorithms above, one can use
benefits of the two languages: let programs to be written in the
syntactic domain, since they are easier to write, then normalise the
programs and let analysis be done on normalised programs, since they
are easier to analyse.  It is an important observation, which can be
generalised to any language possessing canonical forms. Indeed,
compilers use the same approach by transforming programs written in
the flexible surface syntax to a more compact internal
representation. For languages with computational content, such
transformations often improve the performance of the programs.
In this perspective, optimisation of terms can be viewed as
normalisation of terms.

Next section puts this observation to work for EDSLs.

\section{Embedding-By-Normalisation}
\label{sec:EBN}

Embedding is referred to a diverse set of techniques for implementing DSL terms, by
first encoding them as terms in a host language, and then defining
their semantics using the encoded terms. Semantics of DSL terms may be
defined entirely inside the host language by interpreting them in the
host language's runtime system, or partly outside the host language by
compiling code and passing it to an external system.

The key selling points for embedding DSLs are to reuse the machinery
available for a host language, from parser to type checker, and to
integrate with its ecosystem, from editors to run-time system.
EDSLs and embedding techniques that are proven successful in practice,
go beyond traditional sole reuse of syntactic machinery such as parser
and type-checker, and employ the evaluation mechanism of the host
language to optimise the DSL terms
\citep{FELDSPAR,svensson2011obsidian,rompf2012lightweight,
Mainland:2010}.

Briefly put, what these techniques provide is
abstraction-without-guilt: the possibility to define layers of
abstraction in EDSLs, using features available in the host language,
without sacrificing the performance of final produced code.  As
mentioned in the previous section, an optimisation process, such as
the ones used in above techniques, can be viewed as a normalisation
process. So essentially, what the mentioned embedding techniques do is
to perform \textbf{normalisation} of embedded terms \textbf{by}
reusing the \textbf{evaluation} mechanism of the host language.  As
the names suggest, there is a correspondence between such embedding
techniques and NBE begging to be examined:
\begin{center}
optimisation of object by evaluation in host\\
   <--->\\
normalisation of syntax by evaluation in semantic
\end{center}

This section investigates the correspondence, by drawing
parallel between different components of the two sides.
But, before doing so, the class of EDSLs under investigation
should be specified.

\subsection{Normalised EDSLs}
\label{sec:NormalisedEDSLs}
Generally speaking, not every EDSL possess computational content,
e.g., consider DSLs used for data description. On the other hand, a
large and popular class of EDSLs possess some form of computational
content. For the latter, as mentioned earlier, embedding techniques
try to take full advantage of the evaluation process in the host
language to optimise object terms before extracting code from
them. The extracted code is passed, as data, to a back-end, which
either interprets the data by directly calling foreign function
interfaces (e.g., see \citet{MeijerLINQ, accelerate}), or by passing
it to an external compiler (e.g., see
\citet{FELDSPAR,sujeeth2013composition}). This class of EDSLs are
referred to as \emph{normalised EDSLs} in this paper, and they are
distinguished from other EDSLs by the fact that (a) they possess
computational content; (b) the object terms are optimised by using
evaluation in the host language; and (c) they extract code from
optimised object terms and the code is representable as data.

In general, embedding a DSL as a normalised EDSL constitutes of four
components:
\begin{description}
\item [Object Language]
      is the language defining the syntax of the DSL being embedded
\item [Host Language]
      is the language that the DSL is being embedded into
\item [Encoding]
      is the process of defining terms in the object language as a
      specific set of terms in the host language
\item [Code Extraction]
      is the process of deriving object code, as data, from the
      specific set of values (as opposed to general terms) in the host
      language that encode (optimised) object terms
\end{description}

Encoding of object terms as host terms is done in a way that
the resulting values after evaluation of host terms denote optimised
object terms.

For instance, the following is the four components in an embedding
of Chars language:
\begin{itemize}
\item \emph{Object language} is of the following grammar:
\begin{hscode}\SaveRestoreHook
\column{B}{@{}>{\hspre}l<{\hspost}@{}}%
\column{E}{@{}>{\hspre}l<{\hspost}@{}}%
\>[B]{}\Varid{c}\;\Varid{∈}\;\Conid{Char}\;(\Varid{set}\;\Varid{of}\;\Varid{characters}){}\<[E]%
\\
\>[B]{}\Conid{L,M,N}\;\Varid{∈}\;\Conid{Chars}\;\Conid{::=}\;\Varid{ε₀}\;\mid \;\Conid{Chr}\;\Varid{c}\;\mid \;\Conid{M}\;\Varid{∙}\;\Conid{N}{}\<[E]%
\ColumnHook
\end{hscode}\resethooks
\item \emph{Host language} is a pure typed functional language
\item \emph{Encoding} is as follows:\\
\ensuremath{\Varid{ε₀}}    is encoded as the host (nullary) function \ensuremath{eps_f\;\mathrel{=}\;[]} \\
\ensuremath{\Conid{Chr}\;\Varid{c}} is encoded as the host function \ensuremath{chr_f\;\Varid{c}\;\mathrel{=}\;[\mskip1.5mu \;\Varid{c}\;\mskip1.5mu]} \\
\ensuremath{\Conid{M}\;\Varid{∙}\;\Conid{N}} is encoded as the host function \ensuremath{\Conid{M}\;∙_f\;\Conid{N}\;\mathrel{=}\;\Conid{M}\;\plus \;\Conid{N}}
\item \emph{code extraction} is a function from list values
to the datatype (of the $\__d$ indexed constructors) representing
the extracted code
\begin{hscode}\SaveRestoreHook
\column{B}{@{}>{\hspre}l<{\hspost}@{}}%
\column{13}{@{}>{\hspre}l<{\hspost}@{}}%
\column{E}{@{}>{\hspre}l<{\hspost}@{}}%
\>[B]{}\Varid{↓}\;[]\;{}\<[13]%
\>[13]{}\mathrel{=}\;Eps_d{}\<[E]%
\\
\>[B]{}\Varid{↓}\;(\Varid{c}\;\Varid{∷}\;\Varid{cs})\;{}\<[13]%
\>[13]{}\mathrel{=}\;Chr_d\;\Varid{c}\;∙_d\;(\Varid{reify}\;\Varid{cs}){}\<[E]%
\ColumnHook
\end{hscode}\resethooks
\end{itemize}

Users of Chars EDSL write their programs using $\__f$ indexed
functions, and the extracted code, the $\__d$ indexed data, is passed
to back-end of the Chars EDSL.  A simple example of such back-end
would be a function that takes the code and prints the denoted string:

\begin{hscode}\SaveRestoreHook
\column{B}{@{}>{\hspre}l<{\hspost}@{}}%
\column{27}{@{}>{\hspre}l<{\hspost}@{}}%
\column{33}{@{}>{\hspre}l<{\hspost}@{}}%
\column{45}{@{}>{\hspre}l<{\hspost}@{}}%
\column{E}{@{}>{\hspre}l<{\hspost}@{}}%
\>[B]{}\Varid{printChars}\;\mathbin{:}\;Chars_d\;\Varid{→}\;\Conid{IO}\;\Varid{⟨⟩}{}\<[E]%
\\
\>[B]{}\Varid{printChars}\;Eps_d\;{}\<[27]%
\>[27]{}\mathrel{=}\;\Varid{printString}\;\text{\tt \char34 \char34}{}\<[E]%
\\
\>[B]{}\Varid{printChars}\;(Chr_d\;\Varid{c}\;∙_d\;\Conid{N})\;{}\<[27]%
\>[27]{}\mathrel{=}\;\textbf{\text{do}}\;{}\<[33]%
\>[33]{}\Varid{printChar}\;{}\<[45]%
\>[45]{}\Varid{c}\;{}\<[E]%
\\
\>[33]{}\Varid{printChars}\;{}\<[45]%
\>[45]{}\Conid{N}{}\<[E]%
\ColumnHook
\end{hscode}\resethooks

\subsection{Correspondence}
\label{sec:Correspondence}
Comparing embedding structure, explained in Section
\ref{sec:NormalisedEDSLs}, with NBE structure, explained in Section
\ref{sec:NBE}, the correspondence is evident as follows:

\begin{center}
\begin{tabular}{rcl}
NBE              &\ \ <--->\ \ &  EBN \\  \\
Syntactic Domain &\ \ <--->\ \ &  Object Language\\
Semantic  Domain &\ \ <--->\ \ &  Host   Language (a subset of)\\
Evaluation       &\ \ <--->\ \ &  Encoding\\
Reification      &\ \ <--->\ \ &  Code Extraction
\end{tabular}
\end{center}

Viewing embedding through the lens of NBE, one can observe that many
of the smart techniques for encoding object terms as host terms
basically correspond to defining parts of an evaluation process that
maps object terms to values (as opposed to general terms) in a subset
of host language. Once one puts the two sides together, the
correspondence between code extraction and reification in NBE is also
not surprising. Even the name ``reification" has been used by some
embedding experts to refer to the code extraction process (see
\citet{Gill:CACM}).

This paper dubs an embedding process that follows the NBE structure as
Embedding-By-Normalisation, or EBN for short.
Embedding-by-normalisation is to be viewed as a general
theoretical framework to study existing embedding techniques in
practice, and also as a recipe on how to structure implementation of
normalised EDSLs.
EBN builds a bridge between theory and practice: theoretical solutions
in NBE can be used to solve practical problems in embedding, and vice
versa.

There are can be different approaches to perform embedding-by-normalisation.
For instance, provided a back-end to process input code represented as data,
embedding-by-normalisation follows the steps below:
\begin{enumerate}
\item The abstract syntax of the code expected by the back-end
      is identified. Such abstract syntax corresponds to the grammar of
      normal forms.
\item Semantic domain is identified as a subset of the host language.
\item Reification is identified as the process that maps terms in the
      semantic domain to terms in normal form, as usual.
\item Evaluation is identified as programs in the host language
      that map object terms to values in the semantic domain.
\end{enumerate}

Above steps can be reordered. However, important
observation here is that often defining syntax of normal forms and
semantic domain should be prioritised over defining the interface that
the end-users program in, i.e., the syntactic domain.  Syntactic
domain can be seen as the class of host programs that can be
normalised to terms following the grammar of normal forms.

\subsection{Encoding Strategies}
\label{sec:EBN:Encoding}
Due to its correspondence to NBE, EBN is of mathematical nature:
abstract and general. Furthermore, as there are variety of NBE
algorithms, there are variety of corresponding EBN techniques. The
generality and variety make it difficult to propose a concrete
encoding strategy for EBN. The remainder of this section discusses
some general encoding strategies based on the existing techniques
including shallow embedding, final tagless embedding, deep embedding, and
quoted embedding.

\subsubsection{Shallow Embedding}
\label{sec:EBN:Shallow}
Shallow embedding is when an interface formed by a set
of functions in the host is used to represent the syntax of a DSL, and
implementation of the functions as semantics. In shallow embedding,
semantic of the DSL is required to be compositional \citep{Tagless,
Gibbons}.
In EBN, when encoding of object terms follows
shallow embedding, the four components of EBN are as follows:
\begin{description}
\item [Syntactic Domain] is an interface formed by a set of functions
                        (or values) in the host
\item [Semantic Domain] is the result type of above interface
\item [Evaluation] is the overall evaluation of the implementation of
                   the above interface in the host. In this setting,
                   EBN's evaluation process is also required to be
                   compositional, and evaluation of a syntactic term
                   is built up from the evaluation of its sub-terms.
\item [Reification] is a mapping from host values of the semantic
                    domain type to data that implements a subset of
                    syntactic domain interface, i.e., the subset that
                    corresponds to the grammar of normal forms.
\end{description}

The Chars example in Section \ref{sec:NormalisedEDSLs} is EBN with
shallow encoding.

\subsubsection{Final Tagless Embedding}
\label{sec:EBN:Tagless}
Final tagless embedding \citep{Tagless}, which is a specific form of
shallow embedding, is when the shallow interface is parametric over
the semantic type. In Haskell, the parametric interface is defined as
a type-class, where instantiating the type-class defines
semantics. Similar to shallow embedding, in final tagless embedding,
evaluation is required to be compositional.

In EBN, when encoding of object terms follows final tagless embedding,
the four components of EBN are as follows:

\begin{description}
\item [Syntactic Domain] is a type-class (or a similar machinery such as modules)
                        defining syntax in final tagless style
\item [Semantic Domain] is the type that the syntax type-class is
                        instantiated with
\item [Evaluation] is the implementation of an instance of syntax type class.
                   In this setting, EBN's evaluation process is also required
                   to be compositional. An instance of syntax type-class is
                   an algebra for folds over the syntactic language
\item [Reification] is a mapping from host values of the semantic
                    domain type to data that implements a subset of
                    syntactic domain interface, i.e., the subset that
                    corresponds to the grammar of normal forms.
\end{description}

For instance, the following is the four components in EBN of Chars
 language with final tagless encoding:
\begin{itemize}
\item \emph{Syntactic domain} is the following type-class declaration:
\begin{hscode}\SaveRestoreHook
\column{B}{@{}>{\hspre}l<{\hspost}@{}}%
\column{3}{@{}>{\hspre}l<{\hspost}@{}}%
\column{9}{@{}>{\hspre}l<{\hspost}@{}}%
\column{E}{@{}>{\hspre}l<{\hspost}@{}}%
\>[B]{}\textbf{class}\;\Conid{CharsLike}\;\Varid{chars}\;\Keyword{where}{}\<[E]%
\\
\>[B]{}\hsindent{3}{}\<[3]%
\>[3]{}eps_f\;{}\<[9]%
\>[9]{}\mathbin{:}\;\Varid{chars}{}\<[E]%
\\
\>[B]{}\hsindent{3}{}\<[3]%
\>[3]{}chr_f\;{}\<[9]%
\>[9]{}\mathbin{:}\;\Conid{Char}\;\Varid{→}\;\Varid{chars}{}\<[E]%
\\
\>[B]{}\hsindent{3}{}\<[3]%
\>[3]{}(∙_f)\;{}\<[9]%
\>[9]{}\mathbin{:}\;\Varid{chars}\;\Varid{→}\;\Varid{chars}\;\Varid{→}\;\Varid{chars}{}\<[E]%
\ColumnHook
\end{hscode}\resethooks

\item \emph{Semantic domain} is the type \ensuremath{\Conid{List}\;\Conid{Char}} in a functional language with type-classes
\item \emph{Evaluation} is the following type-class instance:
\begin{hscode}\SaveRestoreHook
\column{B}{@{}>{\hspre}l<{\hspost}@{}}%
\column{3}{@{}>{\hspre}l<{\hspost}@{}}%
\column{11}{@{}>{\hspre}l<{\hspost}@{}}%
\column{E}{@{}>{\hspre}l<{\hspost}@{}}%
\>[B]{}\textbf{instance}\;\Conid{CharsLike}\;(\Conid{List}\;\Conid{Char})\;\Keyword{where}{}\<[E]%
\\
\>[B]{}\hsindent{3}{}\<[3]%
\>[3]{}eps_f\;{}\<[11]%
\>[11]{}\mathrel{=}\;[]{}\<[E]%
\\
\>[B]{}\hsindent{3}{}\<[3]%
\>[3]{}chr_f\;\Varid{c}\;{}\<[11]%
\>[11]{}\mathrel{=}\;[\mskip1.5mu \;\Varid{c}\;\mskip1.5mu]{}\<[E]%
\\
\>[B]{}\hsindent{3}{}\<[3]%
\>[3]{}\Varid{m}\;∙_f\;\Varid{n}\;{}\<[11]%
\>[11]{}\mathrel{=}\;\Varid{m}\;\plus \;\Varid{n}{}\<[E]%
\ColumnHook
\end{hscode}\resethooks
\item \emph{Reification} is the following function
\begin{hscode}\SaveRestoreHook
\column{B}{@{}>{\hspre}l<{\hspost}@{}}%
\column{18}{@{}>{\hspre}l<{\hspost}@{}}%
\column{E}{@{}>{\hspre}l<{\hspost}@{}}%
\>[B]{}\Varid{reify}\;\mathbin{:}\;\Conid{List}\;\Conid{Char}\;\Varid{→}\;Chars_d{}\<[E]%
\\
\>[B]{}\Varid{reify}\;[]\;{}\<[18]%
\>[18]{}\mathrel{=}\;Eps_d{}\<[E]%
\\
\>[B]{}\Varid{reify}\;(\Varid{c}\;\Conid{::}\;\Varid{cs})\;{}\<[18]%
\>[18]{}\mathrel{=}\;Chr_d\;\Varid{c}\;∙_d\;(\Varid{reify}\;\Varid{cs}){}\<[E]%
\ColumnHook
\end{hscode}\resethooks
where code is defined as the following algebraic datatype
\begin{hscode}\SaveRestoreHook
\column{B}{@{}>{\hspre}l<{\hspost}@{}}%
\column{14}{@{}>{\hspre}l<{\hspost}@{}}%
\column{17}{@{}>{\hspre}l<{\hspost}@{}}%
\column{E}{@{}>{\hspre}l<{\hspost}@{}}%
\>[B]{}\Keyword{data}\;Chars_d\;{}\<[14]%
\>[14]{}\mathrel{=}\;{}\<[17]%
\>[17]{}Eps_d\;{}\<[E]%
\\
\>[14]{}\mid \;{}\<[17]%
\>[17]{}Chr_d\;\Varid{c}\;{}\<[E]%
\\
\>[14]{}\mid \;{}\<[17]%
\>[17]{}Chars_d\;∙_d\;Chars_d{}\<[E]%
\ColumnHook
\end{hscode}\resethooks
\end{itemize}

\subsubsection{Deep Embedding}
\label{sec:EBN:Deep}
Deep embedding is when datatypes in host are used for representing the
syntax of a DSL, and semantics is defined as functions (programs in
general) over the syntax datatypes.

In EBN, when encoding of object terms follows deep embedding,
the four components of EBN are as follows:

\begin{description}
\item [Syntactic Domain] is a datatype
\item [Semantic Domain] is a type that the syntax datatype is transformed to
\item [Evaluation] is a function from syntax datatype to semantic domain
\item [Reification] is a mapping from host values of the semantic
                    domain type to a datatype describing normal forms
\end{description}

For instance, the following is the four components in EBN of Chars
 language with deep encoding:
\begin{itemize}
\item \emph{Syntactic domain} is the following datatype:
\begin{hscode}\SaveRestoreHook
\column{B}{@{}>{\hspre}l<{\hspost}@{}}%
\column{14}{@{}>{\hspre}l<{\hspost}@{}}%
\column{17}{@{}>{\hspre}l<{\hspost}@{}}%
\column{E}{@{}>{\hspre}l<{\hspost}@{}}%
\>[B]{}\Keyword{data}\;Chars_d\;{}\<[14]%
\>[14]{}\mathrel{=}\;{}\<[17]%
\>[17]{}Eps_d\;{}\<[E]%
\\
\>[14]{}\mid \;{}\<[17]%
\>[17]{}Chr_d\;\Varid{c}\;{}\<[E]%
\\
\>[14]{}\mid \;{}\<[17]%
\>[17]{}Chars_d\;∙_d\;Chars_d{}\<[E]%
\ColumnHook
\end{hscode}\resethooks
\item \emph{Semantic domain} is the type \ensuremath{\Conid{List}\;\Conid{Char}} in a functional language
                             with algebraic datatypes
\item \emph{Evaluation} is the following function:
\begin{hscode}\SaveRestoreHook
\column{B}{@{}>{\hspre}l<{\hspost}@{}}%
\column{7}{@{}>{\hspre}l<{\hspost}@{}}%
\column{17}{@{}>{\hspre}l<{\hspost}@{}}%
\column{E}{@{}>{\hspre}l<{\hspost}@{}}%
\>[B]{}\Varid{eval}\;\mathbin{:}\;Chars_d\;\Varid{→}\;\Conid{List}\;\Conid{Char}{}\<[E]%
\\
\>[B]{}\Varid{eval}\;{}\<[7]%
\>[7]{}Eps_d\;{}\<[17]%
\>[17]{}\mathrel{=}\;[]{}\<[E]%
\\
\>[B]{}\Varid{eval}\;{}\<[7]%
\>[7]{}(Chr_d\;\Varid{c})\;{}\<[17]%
\>[17]{}\mathrel{=}\;[\mskip1.5mu \;\Varid{c}\;\mskip1.5mu]{}\<[E]%
\\
\>[B]{}\Varid{eval}\;{}\<[7]%
\>[7]{}(\Varid{m}\;∙_d\;\Varid{n})\;{}\<[17]%
\>[17]{}\mathrel{=}\;\Varid{m}\;\plus \;\Varid{n}{}\<[E]%
\ColumnHook
\end{hscode}\resethooks
\item \emph{Reification} is the following function
\begin{hscode}\SaveRestoreHook
\column{B}{@{}>{\hspre}l<{\hspost}@{}}%
\column{17}{@{}>{\hspre}l<{\hspost}@{}}%
\column{E}{@{}>{\hspre}l<{\hspost}@{}}%
\>[B]{}\Varid{reify}\;\mathbin{:}\;\Conid{List}\;\Conid{Char}\;\Varid{→}\;Chars_d{}\<[E]%
\\
\>[B]{}\Varid{reify}\;[]\;{}\<[17]%
\>[17]{}\mathrel{=}\;Eps_d{}\<[E]%
\\
\>[B]{}\Varid{reify}\;(\Varid{c}\;\Varid{∷}\;\Varid{cs})\;{}\<[17]%
\>[17]{}\mathrel{=}\;Chr_d\;\Varid{c}\;∙_d\;(\Varid{reify}\;\Varid{cs}){}\<[E]%
\ColumnHook
\end{hscode}\resethooks
\end{itemize}

\subsubsection{Quoted Embedding}
\label{sec:EBN:Quoted}
Quoted embedding \citep{QDSL}, which is a specific form of deep
embedding, is when some form of quotations is used to represent
syntax, and semantics is defined as functions over the unquoted
representation.

In EBN, when encoding of object terms follows quoted embedding,
the four components of EBN are as follows:

\begin{description}
\item [Syntactic Domain] is the type of quoted terms in the host
\item [Semantic Domain] is a type that the unquoted representation of
                        syntactic terms is transformed to
\item [Evaluation] is a function from unquoted representation of
                   syntactic terms to semantic domain
\item [Reification] is a mapping from host values of the semantic
                    domain type to a datatype describing normal forms
\end{description}

For instance, the following is the four components in EBN of Chars
 language with quoted encoding:
\begin{itemize}
\item \emph{Syntactic domain} is \ensuremath{Chars_d}, the resulting type of a
      quasi-quotation denoted as \ensuremath{\Varid{[c|...|]}} for the grammar of Chars language.
\item \emph{Semantic domain} is the type \ensuremath{\Conid{List}\;\Conid{Char}}
      in a functional language with quasi-quotation
\item \emph{Evaluation} is the following function:
\begin{hscode}\SaveRestoreHook
\column{B}{@{}>{\hspre}l<{\hspost}@{}}%
\column{7}{@{}>{\hspre}l<{\hspost}@{}}%
\column{20}{@{}>{\hspre}l<{\hspost}@{}}%
\column{24}{@{}>{\hspre}l<{\hspost}@{}}%
\column{E}{@{}>{\hspre}l<{\hspost}@{}}%
\>[B]{}\Varid{eval}\;\mathbin{:}\;Chars_d\;\Varid{→}\;\Conid{List}\;\Conid{Char}{}\<[E]%
\\
\>[B]{}\Varid{eval}\;{}\<[7]%
\>[7]{}[c|\;\Varid{ε₀}\;{}\<[20]%
\>[20]{}|]\;{}\<[24]%
\>[24]{}\mathrel{=}\;[]{}\<[E]%
\\
\>[B]{}\Varid{eval}\;{}\<[7]%
\>[7]{}[c|\;\Conid{Chr}\;\Varid{\$c}\;{}\<[20]%
\>[20]{}|]\;{}\<[24]%
\>[24]{}\mathrel{=}\;[\mskip1.5mu \;\Varid{c}\;\mskip1.5mu]{}\<[E]%
\\
\>[B]{}\Varid{eval}\;{}\<[7]%
\>[7]{}[c|\;\Varid{\$m}\;\Varid{∙}\;\Varid{\$n}\;{}\<[20]%
\>[20]{}|]\;{}\<[24]%
\>[24]{}\mathrel{=}\;\Varid{m}\;\plus \;\Varid{n}{}\<[E]%
\ColumnHook
\end{hscode}\resethooks
where \ensuremath{\mathbin{\$}} denotes anti-quotation \citep{mainland-quoted}.
\item \emph{Reification} is the following function
\begin{hscode}\SaveRestoreHook
\column{B}{@{}>{\hspre}l<{\hspost}@{}}%
\column{17}{@{}>{\hspre}l<{\hspost}@{}}%
\column{E}{@{}>{\hspre}l<{\hspost}@{}}%
\>[B]{}\Varid{reify}\;\mathbin{:}\;\Conid{List}\;\Conid{Char}\;\Varid{→}\;Chars_d{}\<[E]%
\\
\>[B]{}\Varid{reify}\;[]\;{}\<[17]%
\>[17]{}\mathrel{=}\;[c|\;\Varid{ε₀}\;|]{}\<[E]%
\\
\>[B]{}\Varid{reify}\;(\Varid{c}\;\Varid{∷}\;\Varid{cs})\;{}\<[17]%
\>[17]{}\mathrel{=}\;[c|\;\Conid{Chr}\;\Varid{\$c}\;\Varid{∙}\;\mathbin{\$}\;(\Varid{reify}\;\Varid{cs})\;|]{}\<[E]%
\ColumnHook
\end{hscode}\resethooks
\end{itemize}

\section{Embedding-By-Normalisation, Generically}
\label{sec:Type-Constrained}


Back in 1966, Landin in his landmark paper "The Next 700 Programming Languages"
\citep{Landin1966} argues
that seemingly different programming languages can be seen as
instances of one unified language and that the differences can be
factored as normal libraries for the unified language.  Landin
nominates lambda calculus as the unified language, and shows how to
encode seemingly different language constructs as normal programs in
this language. Since then, Landin's idea has been proven correct over
and over again, evidenced by successful functional programming
languages built based on the very idea (e.g., see the design of
Glasgow Haskell Compiler).

Although Landin's idea was originally expressed in terms of
general-purpose languages, it also applies to domain-specific ones.
Following in his footsteps, this section considers DSLs which can be
expressed using the lambda calculus enriched with primitive values and
operations \citep{Plotkin1975} to express domain-specific constructs.
Not all DSLs can be modelled in this way, and the principles of EBN
applies even outside this model. But this model covers a large and
useful class of DSLs and allows for a parametric presentation of DSLs,
where syntax of a DSL can be identified solely by the signature of the
primitive values and operations.

\subsection{Simple Types and Products}
\label{sec:Basic}
This subsection presents an instantiations of the EBN technique for DSLs
which can be captured by the simply-typed lambda
calculus with product types, parametric over the set of base types,
literals, and the signature of primitive operations. The host language
is assumed to be a pure typed functional language, and the subset the EBN
technique is targeting (i.e., the semantic domain) is identified by a
constraint on the type of host terms.

\subsubsection{Syntactic Domain}
\label{sec:Basic:Syn}
The grammar of types in the object language is standard:
\begin{hscode}\SaveRestoreHook
\column{B}{@{}>{\hspre}l<{\hspost}@{}}%
\column{E}{@{}>{\hspre}l<{\hspost}@{}}%
\>[B]{}\Varid{χ}\;\Varid{∈}\;\Conid{X}\;(\Varid{set}\;\Varid{of}\;\Varid{base}\;\Varid{types}){}\<[E]%
\\
\>[B]{}\Conid{A,B}\;\Varid{∈}\;\underline{\text{Type}}\;\Conid{::=}\;\Varid{χ}\;\mid \;\underline{⟨⟩}\;\mid \;\Conid{A}\;\underline{→}\;\Conid{B}\;\mid \;\Conid{A}\;\underline{×}\;\Conid{B}{}\<[E]%
\ColumnHook
\end{hscode}\resethooks

It is parametric over a set of base types. Besides base types, it
includes unit, function type, and product type.  The types of the
object language are underlined to distinguish it from the ones of the
host language.

The grammar of the terms in the object language is also standard
\citep{Filinski}:

\begin{hscode}\SaveRestoreHook
\column{B}{@{}>{\hspre}l<{\hspost}@{}}%
\column{21}{@{}>{\hspre}l<{\hspost}@{}}%
\column{E}{@{}>{\hspre}l<{\hspost}@{}}%
\>[B]{}\Varid{x}\;\Varid{∈}\;\Conid{Γ}\;(\Varid{set}\;\Varid{of}\;\Varid{variables}){}\<[E]%
\\
\>[B]{}\Varid{ξ}\;\Varid{∈}\;\Conid{Ξ}\;(\Varid{set}\;\Varid{of}\;\Varid{literals}){}\<[E]%
\\
\>[B]{}\Varid{c}\;\Varid{∈}\;\Conid{Σ}\;(\Varid{set}\;\Varid{of}\;\Varid{signature}\;\Varid{of}\;\Varid{primitives}){}\<[E]%
\\
\>[B]{}\Conid{L,M,N}\;\Varid{∈}\;\Conid{Syn}\;\Conid{::=}\;\underline{ξ}\;{}\<[21]%
\>[21]{}\mid \;\Varid{c}\;\overline{M}\;\mid \;\underline{⟨⟩}\;\mid \;\Varid{x}\;\mid \;\underline{λ}\;\Varid{x}\;\underline{→}\;\Conid{N}\;\mid \;\Conid{L}\;\underline{@}\;\Conid{M}\;{}\<[E]%
\\
\>[21]{}\mid \;(\Conid{M}\;\underline{,}\;\Conid{N})\;\mid \;\underline{\Varid{fst}}\;\Conid{L}\;\mid \;\underline{\Varid{snd}}\;\Conid{L}{}\<[E]%
\ColumnHook
\end{hscode}\resethooks

The language, referred to as \ensuremath{\Conid{Syn}}, is parametric over a set of
literals, and signature of primitive operations. Besides literals, and
primitive operations (which are assumed to be fully applied), it
involves unit term, variables, lambda abstraction, application, pairs,
and projections. The terms of the object language are underlined to
distinguish it from the ones of the host language.
The typing rules are the expected ones:

\begin{tabular}{@{}cc@{}}
\\
\ensuremath{\infer{Γ_T\;\Varid{⊢}\;\underline{ξ}\;\mathbin{:}\;\Varid{χ}}{\Varid{ξ}\;\Varid{∈}\;Ξ_T\;\Varid{χ}}}
&
\ensuremath{\infer{Γ_T\;\Varid{⊢}\;\Varid{x}\;\mathbin{:}\;\Conid{A}}{(\Varid{x}\;\mathbin{:}\;\Conid{A})\;\Varid{∈}\;Γ_T}}
\\~\\
\multicolumn{2}{c}{
\ensuremath{\infer{Γ_T\;\Varid{⊢}\;\Varid{c}\;\overline{M}\;\mathbin{:}\;\Conid{B}}{\overline{Γ_T\;\Varid{⊢}\;Mᵢ\;\mathbin{:}\;\Conid{Aᵢ}}\ \ \ \ (\Varid{c}\;\mathbin{:}\;\Varid{[...,Aᵢ,...]}\;\Varid{↦}\;\Conid{B})\;\Varid{∈}\;Σ_T}}}
\\~\\
\ensuremath{\infer{Γ_T\;\Varid{⊢}\;\underline{λ}\;\Varid{x}\;\underline{→}\;\Conid{N}\;\mathbin{:}\;\Conid{A}\;\underline{→}\;\Conid{B}}{Γ_T\;\Varid{,}\;\Varid{x}\;\mathbin{:}\;\Conid{A}\;\Varid{⊢}\;\Conid{N}\;\mathbin{:}\;\Conid{B}}}
&
\ensuremath{\infer{Γ_T\;\Varid{⊢}\;\Conid{L}\;\underline{@}\;\Conid{M}\;\mathbin{:}\;\Conid{B}}{Γ_T\;\Varid{⊢}\;\Conid{L}\;\mathbin{:}\;\Conid{A}\;\underline{→}\;\Conid{B}\ \ \ \ Γ_T\;\Varid{⊢}\;\Conid{M}\;\mathbin{:}\;\Conid{A}}}
\\~\\
\ensuremath{\infer{Γ_T\;\Varid{⊢}\;\underline{⟨⟩}\;\mathbin{:}\;\underline{⟨⟩}}{}}
&
\ensuremath{\infer{Γ_T\;\Varid{⊢}\;(\Conid{M}\;\underline{,}\;\Conid{N})\;\mathbin{:}\;\Conid{A}\;\underline{×}\;\Conid{B}}{Γ_T\;\Varid{⊢}\;\Conid{M}\;\mathbin{:}\;\Conid{A}\ \ \ \ Γ_T\;\Varid{⊢}\;\Conid{N}\;\mathbin{:}\;\Conid{B}}}
\\~\\
\ensuremath{\infer{Γ_T\;\Varid{⊢}\;\underline{\Varid{fst}}\;\Conid{L}\;\mathbin{:}\;\Conid{A}}{Γ_T\;\Varid{⊢}\;\Conid{L}\;\mathbin{:}\;\Conid{A}\;\underline{×}\;\Conid{B}}}
&
\ensuremath{\infer{Γ_T\;\Varid{⊢}\;\underline{\Varid{snd}}\;\Conid{L}\;\mathbin{:}\;\Conid{B}}{Γ_T\;\Varid{⊢}\;\Conid{L}\;\mathbin{:}\;\Conid{A}\;\underline{×}\;\Conid{B}}}
\\~\\
\end{tabular}

In this section, including the other subsections, the EBN technique is
presented in a way that it is independent of encoding strategy: the
underlined terms can be trivially encoded using the standard methods,
such as the ones explained in Section \ref{sec:EBN:Encoding}.  One
possible difficulty might be the treatment of free variables, which
can be trivially addressed by using well-known techniques such as
Higher-Order Abstract Syntax (HOAS) representation \citep{hoas}.
Representation of object terms is assumed be quotient with respect to
alpha-conversion, when new bindings are introduced variables are
assumed to be fresh, and substitutions to be capture avoiding.
Terms in syntactic domain and the normal syntactic terms
are represented in the same way. Though in practice, depending on
encoding, the two may be implemented in different ways. In this paper,
host programs of the type \ensuremath{\Conid{Syn}\;\Conid{A}} refer to both the type of a host
term encoding a term of the type \ensuremath{\Conid{A}} in syntactic domain, and the type
of the extracted code for a normal syntactic term of the type \ensuremath{\Conid{A}}.
The type \ensuremath{\Conid{Syn}\;Γ_T\;\Conid{A}} denotes open \ensuremath{\Conid{Syn}\;\Conid{A}} terms with \ensuremath{Γ_T} being a typing
environment containing the free variables.
Literals in the object language, denoted as \ensuremath{\underline{ξ}}, are a subset of
the ones in the host identified by the set \ensuremath{\Conid{Ξ}}.
To retrieve a corresponding literal value in the host language, the
underline notation is removed.
\ensuremath{Ξ_T} is a mapping from base types in the object language to types in the
host language. \ensuremath{\Varid{ξ}\;\Varid{∈}\;Ξ_T\;\Varid{χ}} denotes a predicate asserting that the
literal \ensuremath{\Varid{ξ}} is an element the type \ensuremath{Ξ_T\;\Varid{χ}} in the host language.
\ensuremath{\Conid{Σ}} is a mapping from the name of a primitive operation to its arity.
Primitive operations are fully applied, a series of arguments, or a
series of typing judgements relating to a primitive is denoted by
overlining.
\ensuremath{Σ_T} is the typing environment for primitive operations, with elements
of the form \ensuremath{\Varid{c}\;\mathbin{:}\;\Varid{[A₀,...,Aₙ]}\;\Varid{↦}\;\Conid{B}} which reads as that the primitive
\ensuremath{\Varid{c}} takes \ensuremath{\Varid{n}\;\Varid{+}\;\Varid{1}} (arity of the primitive) elements of the type \ensuremath{\Conid{Aᵢ}},
for i ranging form \ensuremath{\Varid{0}} to \ensuremath{\Varid{n}}, and returns a value of type \ensuremath{\Conid{B}}.

Now that the syntactic domain has been defined, i.e., the \ensuremath{\Conid{Syn}}
language, it is time to define the semantic domain.

\subsubsection{Semantic Domain}
\label{sec:Basic:Sem}
As the host language is a pure typed functional language, and the
object language being a tiny pure typed functional language itself, a
considerable part of the object language closely mirrors the one of
the host language.  Moreover, the representation of the syntactic
domain itself, is a program in the host language, i.e., a term of the
type \ensuremath{\Conid{Syn}\;\Conid{A}}. This observation is realised by defining semantic domain
as follows:
\begin{hscode}\SaveRestoreHook
\column{B}{@{}>{\hspre}l<{\hspost}@{}}%
\column{E}{@{}>{\hspre}l<{\hspost}@{}}%
\>[B]{}\Conid{Sem}\;\Conid{A}\;\mathrel{=}\;\Varid{∀}\;(\Varid{α}\;\mathbin{:}\;\Conid{Type})\;\Varid{→}\;\Varid{α}\;\Varid{∼}\;\Conid{A}\;\Varid{⇒}\;\Varid{α}{}\<[E]%
\ColumnHook
\end{hscode}\resethooks

\begin{tabular}{cc}
\ensuremath{\infer[\Conid{Synᵣ}]{\Conid{Syn}\;\Conid{A}\;\Varid{∼}\;\Conid{A}}{}}
&
\ensuremath{\infer[\Varid{⟨⟩ᵣ}]{\Varid{⟨⟩}\;\mathord{\sim}\;\underline{⟨⟩}}{}}
\\~\\
\ensuremath{\infer[\Varid{→ᵣ}]{(\Varid{α}\;\Varid{→}\;\Varid{β})\;\Varid{∼}\;(\Conid{A}\;\underline{→}\;\Conid{B})}{\Varid{α}\;\Varid{∼}\;\Conid{A}\ \ \ \ \Varid{β}\;\Varid{∼}\;\Conid{B}}}
&
\ensuremath{\infer[\Varid{×ᵣ}]{(\Varid{α}\;\Varid{×}\;\Varid{β})\;\Varid{∼}\;(\Conid{A}\;\underline{×}\;\Conid{B})}{\Varid{α}\;\Varid{∼}\;\Conid{A}\ \ \ \ \Varid{β}\;\Varid{∼}\;\Conid{B}}}
\\~\\
\end{tabular}

That is, a term of type \ensuremath{\Conid{A}} in the semantic domain is any host term
whose type respects the \ensuremath{\Varid{∼}} relation. The relation \ensuremath{\Varid{∼}} states that (a)
semantic terms of unit, function, and product type correspond to host
terms of similar type, (b) a syntactic term encoded in the host
directly correspond to a semantic term of the same type.
Condition (b) is a distinguishing feature in the definition of
evaluation and semantic domain of NBE. As mentioned in Section
\ref{sec:NBE}, evaluation in NBE is allowed to leave parts of
syntactic terms uninterpreted. An uninterpreted part is referred to
as a residualised part, and the act of leaving a part uninterpreted as
residualising.

\subsubsection{Evaluation}
\label{sec:Basic:Evaluation}
Except for terms of base type, evaluation process is standard:
syntactic terms are mapped to corresponding host terms.  Terms of base
types, however, are residualised.  The definition of evaluation
function is as follows:

\begin{hscode}\SaveRestoreHook
\column{B}{@{}>{\hspre}l<{\hspost}@{}}%
\column{11}{@{}>{\hspre}l<{\hspost}@{}}%
\column{22}{@{}>{\hspre}l<{\hspost}@{}}%
\column{25}{@{}>{\hspre}l<{\hspost}@{}}%
\column{E}{@{}>{\hspre}l<{\hspost}@{}}%
\>[B]{}\Varid{⟦\char95 ⟧}\;\mathbin{:}\;\underline{\text{Type}}\;\Varid{→}\;\Conid{Type}{}\<[E]%
\\[\blanklineskip]%
\>[B]{}\Varid{⟦}\;\Varid{χ}\;{}\<[11]%
\>[11]{}\Varid{⟧}\;\mathrel{=}\;\Conid{Syn}\;\Varid{χ}{}\<[E]%
\\
\>[B]{}\Varid{⟦}\;\underline{⟨⟩}\;{}\<[11]%
\>[11]{}\Varid{⟧}\;\mathrel{=}\;\Varid{⟨⟩}{}\<[E]%
\\
\>[B]{}\Varid{⟦}\;\Conid{A}\;\underline{→}\;\Conid{B}\;{}\<[11]%
\>[11]{}\Varid{⟧}\;\mathrel{=}\;\Varid{⟦}\;\Conid{A}\;\Varid{⟧}\;{}\<[22]%
\>[22]{}\Varid{→}\;{}\<[25]%
\>[25]{}\Varid{⟦}\;\Conid{B}\;\Varid{⟧}{}\<[E]%
\\
\>[B]{}\Varid{⟦}\;\Conid{A}\;\underline{×}\;\Conid{B}\;{}\<[11]%
\>[11]{}\Varid{⟧}\;\mathrel{=}\;\Varid{⟦}\;\Conid{A}\;\Varid{⟧}\;{}\<[22]%
\>[22]{}\Varid{×}\;{}\<[25]%
\>[25]{}\Varid{⟦}\;\Conid{B}\;\Varid{⟧}{}\<[E]%
\ColumnHook
\end{hscode}\resethooks

\begin{hscode}\SaveRestoreHook
\column{B}{@{}>{\hspre}l<{\hspost}@{}}%
\column{14}{@{}>{\hspre}l<{\hspost}@{}}%
\column{23}{@{}>{\hspre}l<{\hspost}@{}}%
\column{E}{@{}>{\hspre}l<{\hspost}@{}}%
\>[B]{}\Varid{⟦\char95 ⟧}\;\mathbin{:}\;\Conid{Syn}\;Γ_T\;\Conid{A}\;\Varid{→}\;\Varid{⟦}\;Σ_T\;\Varid{⟧}\;\Varid{→}\;\Varid{⟦}\;Γ_T\;\Varid{⟧}\;\Varid{→}\;\Varid{⟦}\;\Conid{A}\;\Varid{⟧}{}\<[E]%
\\[\blanklineskip]%
\>[B]{}\Varid{⟦}\;\underline{ξ}\;{}\<[14]%
\>[14]{}\Varid{⟧}\;Σ_V\;Γ_V\;{}\<[23]%
\>[23]{}\mathrel{=}\;\underline{ξ}{}\<[E]%
\\
\>[B]{}\Varid{⟦}\;\Varid{c}\;\overline{M}\;{}\<[14]%
\>[14]{}\Varid{⟧}\;Σ_V\;Γ_V\;{}\<[23]%
\>[23]{}\mathrel{=}\;Σ_V\;\Varid{c}\;\Varid{⟦}\;\overline{M}\;\Varid{⟧}{}\<[E]%
\\
\>[B]{}\Varid{⟦}\;\underline{⟨⟩}\;{}\<[14]%
\>[14]{}\Varid{⟧}\;Σ_V\;Γ_V\;{}\<[23]%
\>[23]{}\mathrel{=}\;\Varid{⟨⟩}{}\<[E]%
\\
\>[B]{}\Varid{⟦}\;\Varid{x}\;{}\<[14]%
\>[14]{}\Varid{⟧}\;Σ_V\;Γ_V\;{}\<[23]%
\>[23]{}\mathrel{=}\;Γ_V\;\Varid{x}{}\<[E]%
\\
\>[B]{}\Varid{⟦}\;\underline{λ}\;\Varid{x}\;\underline{→}\;\Conid{N}\;{}\<[14]%
\>[14]{}\Varid{⟧}\;Σ_V\;Γ_V\;{}\<[23]%
\>[23]{}\mathrel{=}\;\Varid{λ}\;\Varid{y}\;\Varid{→}\;\Varid{⟦}\;\Conid{N}\;\Varid{⟧}\;Σ_V\;(Γ_V\;\Varid{,}\;\Varid{x}\;\Varid{↦}\;\Varid{y}){}\<[E]%
\\
\>[B]{}\Varid{⟦}\;\Conid{L}\;\underline{@}\;\Conid{M}\;{}\<[14]%
\>[14]{}\Varid{⟧}\;Σ_V\;Γ_V\;{}\<[23]%
\>[23]{}\mathrel{=}\;(\Varid{⟦}\;\Conid{L}\;\Varid{⟧}\;Σ_V\;Γ_V)\;(\Varid{⟦}\;\Conid{M}\;\Varid{⟧}\;Σ_V\;Γ_V){}\<[E]%
\\
\>[B]{}\Varid{⟦}\;(\Conid{M}\;\underline{,}\;\Conid{N})\;{}\<[14]%
\>[14]{}\Varid{⟧}\;Σ_V\;Γ_V\;{}\<[23]%
\>[23]{}\mathrel{=}\;(\Varid{⟦}\;\Conid{M}\;\Varid{⟧}\;Σ_V\;Γ_V\;\Varid{,}\;\Varid{⟦}\;\Conid{N}\;\Varid{⟧}\;Σ_V\;Γ_V){}\<[E]%
\\
\>[B]{}\Varid{⟦}\;\underline{\Varid{fst}}\;\Conid{L}\;{}\<[14]%
\>[14]{}\Varid{⟧}\;Σ_V\;Γ_V\;{}\<[23]%
\>[23]{}\mathrel{=}\;\Varid{fst}\;(\Varid{⟦}\;\Conid{L}\;\Varid{⟧}\;Σ_V\;Γ_V){}\<[E]%
\\
\>[B]{}\Varid{⟦}\;\underline{\Varid{snd}}\;\Conid{L}\;{}\<[14]%
\>[14]{}\Varid{⟧}\;Σ_V\;Γ_V\;{}\<[23]%
\>[23]{}\mathrel{=}\;\Varid{snd}\;(\Varid{⟦}\;\Conid{L}\;\Varid{⟧}\;Σ_V\;Γ_V){}\<[E]%
\ColumnHook
\end{hscode}\resethooks

Apart from the input expression, the evaluation function takes two
extra arguments: variable \ensuremath{Σ_V} of type \ensuremath{\Varid{⟦}\;Σ_T\;\Varid{⟧}}, that is the
environment of semantic values corresponding to each primitive operator;
and variable \ensuremath{Γ_V} of type \ensuremath{\Varid{⟦}\;Γ_T\;\Varid{⟧}}, that is the environment of
semantic values corresponding to each free variable.
Following the convention, the semantic bracket notation is overloaded,
and denotes the mapping of different kinds of elements from syntax to
semantic.

\subsubsection{Reification}
\label{sec:Basic:Reify}
The final step is to define the reification function. Reification can
be defined as a function indexed by the relation between syntax and
semantics:

\begin{hscode}\SaveRestoreHook
\column{B}{@{}>{\hspre}l<{\hspost}@{}}%
\column{13}{@{}>{\hspre}l<{\hspost}@{}}%
\column{16}{@{}>{\hspre}l<{\hspost}@{}}%
\column{E}{@{}>{\hspre}l<{\hspost}@{}}%
\>[B]{}\Varid{↓}\;\mathbin{:}\;\Varid{α}\;\Varid{∼}\;\Conid{A}\;\Varid{→}\;\Varid{α}\;\Varid{→}\;\Conid{Syn}\;\Conid{A}{}\<[E]%
\\[\blanklineskip]%
\>[B]{}\Varid{↓}\;\Conid{Synᵣ}\;{}\<[13]%
\>[13]{}\Conid{V}\;{}\<[16]%
\>[16]{}\mathrel{=}\;\Conid{V}{}\<[E]%
\\
\>[B]{}\Varid{↓}\;\Varid{⟨⟩ᵣ}\;{}\<[13]%
\>[13]{}\Conid{V}\;{}\<[16]%
\>[16]{}\mathrel{=}\;\underline{⟨⟩}{}\<[E]%
\\
\>[B]{}\Varid{↓}\;(\Varid{a}\;\Varid{→ᵣ}\;\Varid{b})\;{}\<[13]%
\>[13]{}\Conid{V}\;{}\<[16]%
\>[16]{}\mathrel{=}\;\underline{λ}\;\Varid{x}\;\underline{→}\;\Varid{↓}\;\Varid{b}\;(\Conid{V}\;(\Varid{↑}\;\Varid{a}\;\Varid{x})){}\<[E]%
\\
\>[B]{}\Varid{↓}\;(\Varid{a}\;\Varid{×ᵣ}\;\Varid{b})\;{}\<[13]%
\>[13]{}\Conid{V}\;{}\<[16]%
\>[16]{}\mathrel{=}\;(\Varid{↓}\;\Varid{a}\;(\Varid{fst}\;\Conid{V})\;\underline{,}\;\Varid{↓}\;\Varid{b}\;(\Varid{snd}\;\Conid{V})){}\<[E]%
\ColumnHook
\end{hscode}\resethooks
\begin{hscode}\SaveRestoreHook
\column{B}{@{}>{\hspre}l<{\hspost}@{}}%
\column{13}{@{}>{\hspre}l<{\hspost}@{}}%
\column{16}{@{}>{\hspre}l<{\hspost}@{}}%
\column{E}{@{}>{\hspre}l<{\hspost}@{}}%
\>[B]{}\Varid{↑}\;\mathbin{:}\;\Varid{α}\;\Varid{∼}\;\Conid{A}\;\Varid{→}\;\Conid{Syn}\;\Conid{A}\;\Varid{→}\;\Varid{α}{}\<[E]%
\\[\blanklineskip]%
\>[B]{}\Varid{↑}\;\Conid{Synᵣ}\;{}\<[13]%
\>[13]{}\Conid{M}\;{}\<[16]%
\>[16]{}\mathrel{=}\;\Conid{M}{}\<[E]%
\\
\>[B]{}\Varid{↑}\;\Varid{⟨⟩ᵣ}\;{}\<[13]%
\>[13]{}\Conid{M}\;{}\<[16]%
\>[16]{}\mathrel{=}\;\Varid{⟨⟩}{}\<[E]%
\\
\>[B]{}\Varid{↑}\;(\Varid{a}\;\Varid{→ᵣ}\;\Varid{b})\;{}\<[13]%
\>[13]{}\Conid{M}\;{}\<[16]%
\>[16]{}\mathrel{=}\;\Varid{λ}\;\Varid{x}\;\Varid{→}\;\Varid{↑}\;\Varid{b}\;(\Conid{M}\;\underline{@}\;(\Varid{↓}\;\Varid{a}\;\Varid{x})){}\<[E]%
\\
\>[B]{}\Varid{↑}\;(\Varid{a}\;\Varid{×ᵣ}\;\Varid{b})\;{}\<[13]%
\>[13]{}\Conid{M}\;{}\<[16]%
\>[16]{}\mathrel{=}\;(\Varid{↑}\;\Varid{a}\;(\underline{\Varid{fst}}\;\Conid{M})\;\Varid{,}\;\Varid{↑}\;\Varid{b}\;(\underline{\Varid{snd}}\;\Conid{M})){}\<[E]%
\ColumnHook
\end{hscode}\resethooks

Above definition is similar to some classic NBE algorithms such as
\citet{Berger} and \citet{TDPE}. Essentially, what above does is a
form of η-expansion in two levels: object language and host language
\citep{TDPE}.  The reification function \ensuremath{\Varid{↓}} is mutually defined with
the function \ensuremath{\Varid{↑}} referred to as the reflection function.

Embedding a term by the EBN algorithm defined in this subsection
results in a code for the corresponding term in η-long β-normal form.
However, the normal form is not extensional, in that two normal terms may be
equivalent but syntactically distinct (see Section \ref{sec:Richer}).

\subsection{Sums}
\label{sec:Sums}
This section extends the EBN technique of the previous subsection,
to support DSL programs involving sum types, such as conditional
expressions.

\subsubsection{Syntactic Domain}
\label{sec:Sums:Syn}
The grammar of the syntactic domain in Section \ref{sec:Basic:Syn} is
extended as follows:
\begin{hscode}\SaveRestoreHook
\column{B}{@{}>{\hspre}l<{\hspost}@{}}%
\column{E}{@{}>{\hspre}l<{\hspost}@{}}%
\>[B]{}\Conid{A,B}\;\Conid{::=}\;\Varid{...}\;\mid \;\Conid{A}\;\underline{+}\;\Conid{B}{}\<[E]%
\ColumnHook
\end{hscode}\resethooks
\begin{hscode}\SaveRestoreHook
\column{B}{@{}>{\hspre}l<{\hspost}@{}}%
\column{12}{@{}>{\hspre}l<{\hspost}@{}}%
\column{E}{@{}>{\hspre}l<{\hspost}@{}}%
\>[B]{}\Conid{L,M,N}\;\Conid{::=}\;{}\<[12]%
\>[12]{}\Varid{...}\;\mid \;\underline{\Varid{inr}}\;\Conid{M}\;\mid \;\underline{\Varid{inr}}\;\Conid{N}\;\mid \;\underline{\Varid{case}}\;\Conid{L}\;\Conid{M}\;\Conid{N}{}\<[E]%
\ColumnHook
\end{hscode}\resethooks
\begin{tabular}{@{}cc@{}}
\\
...
\\~\\
\ensuremath{\infer{Γ_T\;\Varid{⊢}\;\underline{\Varid{inl}}\;\Conid{M}\;\mathbin{:}\;\Conid{A}\;\underline{+}\;\Conid{B}}{Γ_T\;\Varid{⊢}\;\Conid{M}\;\mathbin{:}\;\Conid{A}}}
&
\ensuremath{\infer{Γ_T\;\Varid{⊢}\;\underline{\Varid{inr}}\;\Conid{N}\;\mathbin{:}\;\Conid{A}\;\underline{+}\;\Conid{B}}{Γ_T\;\Varid{⊢}\;\Conid{N}\;\mathbin{:}\;\Conid{B}}}
\\~\\
\multicolumn{2}{c}{
\ensuremath{\infer{Γ_T\;\Varid{⊢}\;\underline{\Varid{case}}\;\Conid{L}\;\Conid{M}\;\Conid{N}\;\mathbin{:}\;\Conid{C}}{Γ_T\;\Varid{⊢}\;\Conid{L}\;\mathbin{:}\;\Conid{A}\;\underline{+}\;\Conid{B}\ \ \ Γ_T\;\Varid{⊢}\;\Conid{M}\;\mathbin{:}\;\Conid{A}\;\underline{→}\;\Conid{C}\ \ \ Γ_T\;\Varid{⊢}\;\Conid{N}\;\mathbin{:}\;\Conid{B}\;\underline{→}\;\Conid{C}}}}
\\~\\
\end{tabular}

The extensions are standard: sum types, left injection, right
injection, and case expression. To simplify the presentation, branches
of the case expression \ensuremath{\underline{\Varid{case}}\;\Conid{L}\;\Conid{M}\;\Conid{N}} (i.e., M and N) are standard terms
of function type, as opposed to a specific built-in language
constructs with bindings.  It follows Alonzo Church's original idea
that all variable bindings in syntax can be done via bindings in
lambda abstractions.

\subsubsection{Semantic Domain}
\label{sec:Sums:Sem}
To support sum types, it is not enough to simply add a clause to the
relation \ensuremath{\Varid{∼}} of Section \ref{sec:Basic:Sem} relating sum types in the
host to the ones in the object. Treating sum types has been a
challenging problem in NBE and embedding. Essentially, to reify a
semantic term of the type \ensuremath{(\Conid{Syn}\;\Conid{A}\;\Varid{+}\;\Conid{Syn}\;\Conid{B})\;\Varid{→}\;\Conid{Syn}\;\Conid{C}}, following the
same symmetric style of reify-reflect process in Section
\ref{sec:Basic:Reify}, one needs (due to contravariance of function
type) to convert a syntactic term of the type \ensuremath{\Conid{Syn}\;(\Conid{A}\;\underline{+}\;\Conid{B})} to a
semantic term of the type \ensuremath{\Conid{Syn}\;\Conid{A}\;\Varid{+}\;\Conid{Syn}\;\Conid{B}}. The conversion of the type
\ensuremath{\Conid{Syn}\;(\Conid{A}\;\underline{+}\;\Conid{B})\;\Varid{→}\;\Conid{Syn}\;\Conid{A}\;\Varid{+}\;\Conid{Syn}\;\Conid{B}} is problematic, since there is no way
to destruct a term of the type \ensuremath{\Conid{Syn}} and remain in the semantic
domain; the output type of the function is a term in the semantic
domain, while destructing a sum type in syntactic domain demands a
continuation in the syntactic domain. For a more detailed account of
the reification problem for sum types, refer to
\citet{QDSL,Gill:CACM,svenningsson:combiningJournal}.
In the context of type-directed partial evaluation, a NBE based
technique, \citet{TDPE} proposed an elegant solution to the problem of
reification of sums, using composable continuations (delimited
continuations) based on shift and reset \citep{Delimited}. This paper
employs Danvy's solution.

Delimited continuations are effect-full constructs. To model them in
the pure and typed setting of the host language, this paper uses the
standard monadic semantic (e.g., see \citep{Atkey,Dyvbig,Wadler}).
The \ensuremath{\Varid{∼}} relation from Section \ref{sec:Basic:Sem} is updated as
follows:
\begin{hscode}\SaveRestoreHook
\column{B}{@{}>{\hspre}l<{\hspost}@{}}%
\column{E}{@{}>{\hspre}l<{\hspost}@{}}%
\>[B]{}\Varid{...}{}\<[E]%
\ColumnHook
\end{hscode}\resethooks
\begin{tabular}{cc}
\ensuremath{\infer[\Varid{→ᵣ}]{(\Varid{α}\;\Varid{↝}\;\Varid{β})\;\Varid{∼}\;(\Conid{A}\;\underline{→}\;\Conid{B})}{\Varid{α}\;\Varid{∼}\;\Conid{A}\ \ \ \ \Varid{β}\;\Varid{∼}\;\Conid{B}}}
&
\ensuremath{\infer[\Varid{+ᵣ}]{(\Varid{α}\;\Varid{+}\;\Varid{β})\;\Varid{∼}\;(\Conid{A}\;\underline{+}\;\Conid{B})}{\Varid{α}\;\Varid{∼}\;\Conid{A}\ \ \ \ \Varid{β}\;\Varid{∼}\;\Conid{B}}}
\\~\\
\end{tabular}

where \ensuremath{\Varid{↝}} denotes type of monadic functions, i.e., effect-full
functions modelled in the mentioned standard monadic semantic.

One subtle, yet important factor in play here is the perspective that
EBN offers: Danvy's elegant use of shift and reset is not a mere
technical solution (even if it may seem like so when used in an untyped
impure language); through the lens of NBE/EBN, it can be seen as a
change of semantic domain to a monadic one, where the use of shift and
reset are the resulting consequences.

\subsubsection{Evaluation}
\label{sec:Sums:Evaluation}

Since we will use a monad for delimited continuations the semantic domain
will be updated so that functions in the object language are now mapped to
monadic functions in the host language. As anticipated, we also add sums,
which maps to sums in the host language.

\begin{hscode}\SaveRestoreHook
\column{B}{@{}>{\hspre}l<{\hspost}@{}}%
\column{11}{@{}>{\hspre}l<{\hspost}@{}}%
\column{22}{@{}>{\hspre}l<{\hspost}@{}}%
\column{25}{@{}>{\hspre}l<{\hspost}@{}}%
\column{E}{@{}>{\hspre}l<{\hspost}@{}}%
\>[B]{}\Varid{...}{}\<[E]%
\\
\>[B]{}\Varid{⟦}\;\Conid{A}\;\underline{→}\;\Conid{B}\;{}\<[11]%
\>[11]{}\Varid{⟧}\;\mathrel{=}\;\Varid{⟦}\;\Conid{A}\;\Varid{⟧}\;{}\<[22]%
\>[22]{}\Varid{↝}\;{}\<[25]%
\>[25]{}\Varid{⟦}\;\Conid{B}\;\Varid{⟧}{}\<[E]%
\\
\>[B]{}\Varid{⟦}\;\Conid{A}\;\underline{+}\;\Conid{B}\;{}\<[11]%
\>[11]{}\Varid{⟧}\;\mathrel{=}\;\Varid{⟦}\;\Conid{A}\;\Varid{⟧}\;{}\<[22]%
\>[22]{}\Varid{+}\;{}\<[25]%
\>[25]{}\Varid{⟦}\;\Conid{B}\;\Varid{⟧}{}\<[E]%
\ColumnHook
\end{hscode}\resethooks

The evaluator now needs to be updated to reflect the fact that the semantic
domain uses monadic functions. All of the cases from the evaluator Section
\ref{sec:Basic:Evaluation} have been updated to lift the result into a monad.
\begin{hscode}\SaveRestoreHook
\column{B}{@{}>{\hspre}l<{\hspost}@{}}%
\column{15}{@{}>{\hspre}l<{\hspost}@{}}%
\column{16}{@{}>{\hspre}l<{\hspost}@{}}%
\column{24}{@{}>{\hspre}l<{\hspost}@{}}%
\column{25}{@{}>{\hspre}l<{\hspost}@{}}%
\column{40}{@{}>{\hspre}l<{\hspost}@{}}%
\column{E}{@{}>{\hspre}l<{\hspost}@{}}%
\>[B]{}\Varid{⟦\char95 ⟧}\;\mathbin{:}\;\Conid{Syn}\;Γ_T\;\Conid{A}\;\Varid{→}\;\Varid{⟦}\;Σ_T\;\Varid{⟧}\;\Varid{→}\;\Varid{⟦}\;Γ_T\;\Varid{⟧}\;\Varid{→}\;\Varid{⟦}\;\Conid{A}\;\Varid{⟧}{}\<[E]%
\\[\blanklineskip]%
\>[B]{}\Varid{⟦}\;\underline{ξ}\;{}\<[15]%
\>[15]{}\Varid{⟧}\;Σ_V\;Γ_V\;{}\<[24]%
\>[24]{}\mathrel{=}\;\Varid{⦇}\;\underline{ξ}\;\Varid{⦈}{}\<[E]%
\\
\>[B]{}\Varid{⟦}\;\Varid{c}\;\overline{M}\;{}\<[15]%
\>[15]{}\Varid{⟧}\;Σ_V\;Γ_V\;{}\<[24]%
\>[24]{}\mathrel{=}\;Σ_V\;\Varid{c}\;\Varid{⟦}\;\overline{M}\;\Varid{⟧}{}\<[E]%
\\
\>[B]{}\Varid{⟦}\;\underline{⟨⟩}\;{}\<[15]%
\>[15]{}\Varid{⟧}\;Σ_V\;Γ_V\;{}\<[24]%
\>[24]{}\mathrel{=}\;\Varid{⦇}\;\Varid{⟨⟩}\;\Varid{⦈}{}\<[E]%
\\
\>[B]{}\Varid{⟦}\;\Varid{x}\;{}\<[15]%
\>[15]{}\Varid{⟧}\;Σ_V\;Γ_V\;{}\<[24]%
\>[24]{}\mathrel{=}\;Γ_V\;\Varid{x}{}\<[E]%
\\
\>[B]{}\Varid{⟦}\;\underline{λ}\;\Varid{x}\;\underline{→}\;\Conid{N}\;{}\<[15]%
\>[15]{}\Varid{⟧}\;Σ_V\;Γ_V\;{}\<[24]%
\>[24]{}\mathrel{=}\;\Varid{⦇}\;\Varid{λ}\;\Varid{y}\;\Varid{→}\;\Varid{⟦}\;\Conid{N}\;\Varid{⟧}\;Σ_V\;(Γ_V\;\Varid{,}\;\Varid{x}\;\Varid{↦}\;\Varid{⦇}\;\Varid{y}\;\Varid{⦈})\;\Varid{⦈}{}\<[E]%
\\
\>[B]{}\Varid{⟦}\;\Conid{L}\;\underline{@}\;\Conid{M}\;{}\<[15]%
\>[15]{}\Varid{⟧}\;Σ_V\;Γ_V\;{}\<[24]%
\>[24]{}\mathrel{=}\;\Varid{join}\;\Varid{⦇}\;(\Varid{⟦}\;\Conid{L}\;\Varid{⟧}\;Σ_V\;Γ_V)\;(\Varid{⟦}\;\Conid{M}\;\Varid{⟧}\;Σ_V\;Γ_V)\;\Varid{⦈}{}\<[E]%
\\
\>[B]{}\Varid{⟦}\;(\Conid{M}\;\underline{,}\;\Conid{N})\;{}\<[15]%
\>[15]{}\Varid{⟧}\;Σ_V\;Γ_V\;{}\<[24]%
\>[24]{}\mathrel{=}\;\Varid{⦇}\;(\Varid{⟦}\;\Conid{M}\;\Varid{⟧}\;Σ_V\;Γ_V\;\Varid{,}\;\Varid{⟦}\;\Conid{N}\;\Varid{⟧}\;Σ_V\;Γ_V)\;\Varid{⦈}{}\<[E]%
\\
\>[B]{}\Varid{⟦}\;\underline{\Varid{fst}}\;\Conid{L}\;{}\<[15]%
\>[15]{}\Varid{⟧}\;Σ_V\;Γ_V\;{}\<[24]%
\>[24]{}\mathrel{=}\;\Varid{⦇}\;\Varid{fst}\;(\Varid{⟦}\;\Conid{L}\;\Varid{⟧}\;Σ_V\;Γ_V)\;\Varid{⦈}{}\<[E]%
\\
\>[B]{}\Varid{⟦}\;\underline{\Varid{snd}}\;\Conid{L}\;{}\<[15]%
\>[15]{}\Varid{⟧}\;Σ_V\;Γ_V\;{}\<[24]%
\>[24]{}\mathrel{=}\;\Varid{⦇}\;\Varid{snd}\;(\Varid{⟦}\;\Conid{L}\;\Varid{⟧}\;Σ_V\;Γ_V)\;\Varid{⦈}{}\<[E]%
\\
\>[B]{}\Varid{⟦}\;\underline{\Varid{inl}}\;\Conid{M}\;{}\<[15]%
\>[15]{}\Varid{⟧}\;Σ_V\;Γ_V\;{}\<[24]%
\>[24]{}\mathrel{=}\;\Varid{⦇}\;\Varid{inl}\;(\Varid{⟦}\;\Conid{M}\;\Varid{⟧}\;Σ_V\;Γ_V)\;\Varid{⦈}{}\<[E]%
\\
\>[B]{}\Varid{⟦}\;\underline{\Varid{inr}}\;\Conid{N}\;{}\<[15]%
\>[15]{}\Varid{⟧}\;Σ_V\;Γ_V\;{}\<[24]%
\>[24]{}\mathrel{=}\;\Varid{⦇}\;\Varid{inr}\;(\Varid{⟦}\;\Conid{N}\;\Varid{⟧}\;Σ_V\;Γ_V)\;\Varid{⦈}{}\<[E]%
\\
\>[B]{}\Varid{⟦}\;\underline{\Varid{case}}\;\Conid{L}\;\Conid{M}\;\Conid{N}\;{}\<[16]%
\>[16]{}\Varid{⟧}\;Σ_V\;Γ_V\;{}\<[25]%
\>[25]{}\mathrel{=}\;\Varid{join}\;\Varid{⦇}\;\Varid{case}\;{}\<[40]%
\>[40]{}(\Varid{⟦}\;\Conid{L}\;\Varid{⟧}\;Σ_V\;Γ_V)\;{}\<[E]%
\\
\>[40]{}(\Varid{⟦}\;\Conid{M}\;\Varid{⟧}\;Σ_V\;Γ_V)\;{}\<[E]%
\\
\>[40]{}(\Varid{⟦}\;\Conid{N}\;\Varid{⟧}\;Σ_V\;Γ_V)\;\Varid{⦈}{}\<[E]%
\ColumnHook
\end{hscode}\resethooks
For clarity of presentation, applicative bracket notation
\citep{Applicative} is used in above (denoted as \ensuremath{\Varid{⦇}\;\Varid{...}\;\Varid{⦈}} .
An applicative bracket notation \ensuremath{\Varid{⦇}\;\Conid{L}\;\Conid{M₀}\;\Varid{...}\;\Conid{Mₙ}\;\Varid{⦈}} is a mere syntactic sugar
equivalent to the following term using monadic do notation:
\begin{hscode}\SaveRestoreHook
\column{B}{@{}>{\hspre}l<{\hspost}@{}}%
\column{4}{@{}>{\hspre}l<{\hspost}@{}}%
\column{8}{@{}>{\hspre}l<{\hspost}@{}}%
\column{12}{@{}>{\hspre}l<{\hspost}@{}}%
\column{E}{@{}>{\hspre}l<{\hspost}@{}}%
\>[4]{}\textbf{\text{do}}\;{}\<[8]%
\>[8]{}\Varid{x₀}\;{}\<[12]%
\>[12]{}\Varid{←}\;\Conid{M₀}\;{}\<[E]%
\\
\>[8]{}\Varid{...}\;{}\<[E]%
\\
\>[8]{}\Varid{xₙ}\;{}\<[12]%
\>[12]{}\Varid{←}\;\Conid{Mₙ}\;{}\<[E]%
\\
\>[8]{}\Varid{return}\;(\Conid{L}\;\Varid{x₀}\;\Varid{...}\;\Varid{xₙ}){}\<[E]%
\ColumnHook
\end{hscode}\resethooks
The function \ensuremath{\Varid{join}} is a the well-known monad join function, commonly
used for flattening nested monadic structures.


\subsubsection{Reification}
\label{sec:Sums:Reification}
To adopt Danvy's solution, the Reification process of
\ref{sec:Basic:Reify} is updated as follows:

\begin{hscode}\SaveRestoreHook
\column{B}{@{}>{\hspre}l<{\hspost}@{}}%
\column{13}{@{}>{\hspre}l<{\hspost}@{}}%
\column{16}{@{}>{\hspre}l<{\hspost}@{}}%
\column{26}{@{}>{\hspre}l<{\hspost}@{}}%
\column{E}{@{}>{\hspre}l<{\hspost}@{}}%
\>[B]{}\Varid{...}{}\<[E]%
\\
\>[B]{}\Varid{↓}\;(\Varid{a}\;\Varid{→ᵣ}\;\Varid{b})\;{}\<[13]%
\>[13]{}\Conid{V}\;{}\<[16]%
\>[16]{}\mathrel{=}\;\Varid{λ}\;\Varid{x}\;\Varid{→}\;\Varid{reset}\;\Varid{⦇}\;\Varid{↓}\;\Varid{b}\;(\Varid{join}\;\Varid{⦇}\;\Conid{V}\;(\Varid{↑}\;\Varid{a}\;\Varid{x})\;\Varid{⦈})\;\Varid{⦈}{}\<[E]%
\\
\>[B]{}\Varid{↓}\;(\Varid{a}\;\Varid{+ᵣ}\;\Varid{b})\;{}\<[13]%
\>[13]{}\Conid{V}\;{}\<[16]%
\>[16]{}\mathrel{=}\;\Varid{case}\;\Conid{V}\;{}\<[26]%
\>[26]{}(\Varid{λ}\;\Varid{x}\;\Varid{→}\;\underline{\Varid{inl}}\;(\Varid{↓}\;\Varid{a}\;\Varid{x}))\;{}\<[E]%
\\
\>[26]{}(\Varid{λ}\;\Varid{y}\;\Varid{→}\;\underline{\Varid{inr}}\;(\Varid{↓}\;\Varid{b}\;\Varid{y})){}\<[E]%
\ColumnHook
\end{hscode}\resethooks
\begin{hscode}\SaveRestoreHook
\column{B}{@{}>{\hspre}l<{\hspost}@{}}%
\column{13}{@{}>{\hspre}l<{\hspost}@{}}%
\column{16}{@{}>{\hspre}l<{\hspost}@{}}%
\column{19}{@{}>{\hspre}l<{\hspost}@{}}%
\column{26}{@{}>{\hspre}l<{\hspost}@{}}%
\column{29}{@{}>{\hspre}l<{\hspost}@{}}%
\column{E}{@{}>{\hspre}l<{\hspost}@{}}%
\>[B]{}\Varid{↑}\;\mathbin{:}\;\Varid{α}\;\Varid{∼}\;\Conid{A}\;\Varid{→}\;\Conid{Syn}\;\Conid{A}\;\Varid{↝}\;\Varid{α}{}\<[E]%
\\[\blanklineskip]%
\>[B]{}\Varid{↑}\;\Conid{Synᵣ}\;{}\<[13]%
\>[13]{}\Conid{M}\;{}\<[16]%
\>[16]{}\mathrel{=}\;\Varid{⦇}\;\Conid{M}\;\Varid{⦈}{}\<[E]%
\\
\>[B]{}\Varid{↑}\;\Varid{⟨⟩ᵣ}\;{}\<[13]%
\>[13]{}\Conid{M}\;{}\<[16]%
\>[16]{}\mathrel{=}\;\Varid{⦇}\;\Varid{⟨⟩}\;\Varid{⦈}{}\<[E]%
\\
\>[B]{}\Varid{↑}\;(\Varid{a}\;\Varid{→ᵣ}\;\Varid{b})\;{}\<[13]%
\>[13]{}\Conid{M}\;{}\<[16]%
\>[16]{}\mathrel{=}\;\Varid{⦇}\;\Varid{λ}\;\Varid{x}\;\Varid{→}\;\Varid{↑}\;\Varid{b}\;(\Conid{M}\;\underline{@}\;(\Varid{↓}\;\Varid{a}\;\Varid{x}))\;\Varid{⦈}{}\<[E]%
\\
\>[B]{}\Varid{↑}\;(\Varid{a}\;\Varid{×ᵣ}\;\Varid{b})\;{}\<[13]%
\>[13]{}\Conid{M}\;{}\<[16]%
\>[16]{}\mathrel{=}\;\Varid{⦇}\;(\Varid{↑}\;\Varid{a}\;(\underline{\Varid{fst}}\;\Conid{M})\;\Varid{,}\;\Varid{↑}\;\Varid{b}\;(\underline{\Varid{snd}}\;\Conid{M}))\;\Varid{⦈}{}\<[E]%
\\
\>[B]{}\Varid{↑}\;(\Varid{a}\;\Varid{+ᵣ}\;\Varid{b})\;{}\<[13]%
\>[13]{}\Conid{M}\;{}\<[16]%
\>[16]{}\mathrel{=}\;{}\<[19]%
\>[19]{}\Varid{shift}\;{}\<[26]%
\>[26]{}(\Varid{λ}\;\Varid{k}\;\Varid{→}{}\<[E]%
\\
\>[19]{}\underline{\Varid{case}}\;\Conid{M}\;{}\<[29]%
\>[29]{}(\Varid{λ}\;\Varid{x}\;\Varid{→}\;\Varid{reset}\;\Varid{⦇}\;(\Varid{k}\;\Varid{∘}\;\Varid{inl})\;(\Varid{↑}\;\Varid{a}\;\Varid{x})\;\Varid{⦈})\;{}\<[E]%
\\
\>[29]{}(\Varid{λ}\;\Varid{y}\;\Varid{→}\;\Varid{reset}\;\Varid{⦇}\;(\Varid{k}\;\Varid{∘}\;\Varid{inr})\;(\Varid{↑}\;\Varid{b}\;\Varid{y})\;\Varid{⦈})){}\<[E]%
\ColumnHook
\end{hscode}\resethooks

Except for the necessary monadic liftings, the nontrivial change is
the clause related to sum types in the reflection function.  Instead
of destructing \ensuremath{\Conid{M}} directly, which results in another term of \ensuremath{\Conid{Syn}}
type, first it asks for a continuation \ensuremath{\Varid{k}} that given a semantic value
of sum type, produces a syntactic term, and then uses this
continuation for constructing the syntactic continuations needed for
destructing \ensuremath{\Conid{M}}. Reader interested in more details, may try to follow
above algorithm step-by-step to reify the semantic term \ensuremath{\Varid{λ}\;\Varid{x}\;\Varid{→}\;\underline{⟨⟩}} of
the type \ensuremath{(\Conid{Syn}\;\underline{⟨⟩}\;\Varid{+}\;\Conid{Syn}\;\underline{⟨⟩})\;\Varid{→}\;\Conid{Syn}\;\underline{⟨⟩}}, or the term \ensuremath{\Varid{λ}\;\Varid{x}\;\Varid{→}\;\Varid{fst}\;\Varid{x}} of
the same type, and consult \citet{TDPE}, if needed.

\subsection{Smart Primitives}
\label{sec:Smart}
So far, syntactic terms of base types have been residualised: they are
treated as uninterpreted entities. This subsection proposes an
alternative semantic domain, so that some of the primitives can be
mapped to their corresponding host programs and terms with syntactic
terms with base types get partially normalised.

\subsubsection{Syntactic Domain}
\label{sec:Smart:Syn}
Syntactic domain in this subsection is the same as the one in
Section \ref{sec:Sums:Syn}.

\subsubsection{Semantic Domain}
\label{sec:Smart:Sem}
The type relation \ensuremath{\Varid{∼}} in \ref{sec:Sums:Sem} does not consider
convertibility of semantic values in the host language: given object
type A, if value \ensuremath{\Conid{V}} in the host is convertible, by a function in the
host, to another value \ensuremath{\Conid{W}} which respects \ensuremath{\Varid{∼}\;\Conid{A}}, \ensuremath{\Conid{V}} also respects \ensuremath{\Varid{∼}\;\Conid{A}}. The following is a generalisation of the type relation \ensuremath{\Varid{∼}} in
\ref{sec:Sums:Sem}, based on this observation:

\begin{tabular}{@{}c@{}c@{}}
\\
\ensuremath{\infer[\Conid{Synᵣ}]{\Conid{Syn}\;\Conid{A}\;∼^p\;\Conid{A}}{}}
&
\ensuremath{\infer[\Varid{⟨⟩ᵣ}]{\Varid{⟨⟩}\;∼^p\;\underline{⟨⟩}}{}}
\\~\\
\multicolumn{2}{c}{
\ensuremath{\infer[\Varid{→ᵣ}]{(\Varid{α}\;\Varid{↝}\;\Varid{β})\;∼^p\;(\Conid{A}\;\underline{→}\;\Conid{B})}{(\Varid{α}\;∼^{¬p}\;\Conid{A})\ \ \ \ (\Varid{β}\;∼^p\;\Conid{B})}}
}
\\~\\
\ensuremath{\infer[\Varid{×ᵣ}]{(\Varid{α}\;\Varid{×}\;\Varid{β})\;∼^p\;(\Conid{A}\;\underline{×}\;\Conid{B})}{(\Varid{α}\;∼^p\;\Conid{A})\ \ \ \ (\Varid{β}\;∼^p\;\Conid{B})}}
&
\ensuremath{\infer[\Varid{+ᵣ}]{(\Varid{α}\;\Varid{+}\;\Varid{β})\;∼^p\;(\Conid{A}\;\underline{+}\;\Conid{B})}{(\Varid{α}\;∼^p\;\Conid{A})\ \ \ \ (\Varid{β}\;∼^p\;\Conid{B})}}
\\~\\
\ensuremath{\infer[\Varid{⇑⁻ᵣ}]{\Varid{β}\;\Varid{∼⁻}\;\Conid{A}}{(\Varid{α}\;\Varid{∼⁻}\;\Conid{A})\ \ \ \ (\Varid{∃}\;\Varid{f}\;\mathbin{:}\;\Varid{α}\;\Varid{→}\;\Varid{β})}}
&
\ensuremath{\infer[\Varid{⇑⁺ᵣ}]{\Varid{α}\;\Varid{∼⁺}\;\Conid{A}}{(\Varid{β}\;\Varid{∼⁺}\;\Conid{A})\ \ \ \ (\Varid{∃}\;\Varid{f}\;\mathbin{:}\;\Varid{α}\;\Varid{→}\;\Varid{β})}}
\\~\\
\end{tabular}

Notice that the polarity, or variance, of a type should be tracked as
the type of conversion functions changes direction due to
contravariance of arguments in function types. For instance, consider
function \ensuremath{\Varid{f}} converting \ensuremath{\Conid{V}\;\mathbin{:}\;\Conid{Syn}\;\Conid{A}\;\Varid{→}\;\Conid{Syn}\;\Conid{B}} to \ensuremath{\Varid{f}\;\Conid{V}\;\mathbin{:}\;(\Conid{Syn}\;\Conid{A}\;\Varid{,}\;\Conid{Syn}\;\Conid{A})\;\Varid{→}\;\Conid{Syn}\;\Conid{B}}. One should provide a function \ensuremath{\Varid{g}} of type \ensuremath{(\Conid{Syn}\;\Conid{A}\;\Varid{,}\;\Conid{Syn}\;\Conid{A})\;\Varid{→}\;\Conid{Syn}\;\Conid{A}}, rather than \ensuremath{\Conid{Syn}\;\Conid{A}\;\Varid{→}\;(\Conid{Syn}\;\Conid{A}\;\Varid{,}\;\Conid{Syn}\;\Conid{A})}, so that \ensuremath{\Varid{f}\;\Conid{V}\;\mathrel{=}\;\Varid{λ}\;\Varid{x}\;\Varid{→}\;\Conid{V}\;(\Varid{g}\;\Varid{x})}.

\subsubsection{Evaluation}
\label{sec:Smart:Evaluation}
Evaluation is similar to the one in Section \ref{sec:Sums:Evaluation},
except that here base types can be either residual syntactic terms, as
before, or the corresponding values in the semantic domain:
\begin{hscode}\SaveRestoreHook
\column{B}{@{}>{\hspre}l<{\hspost}@{}}%
\column{11}{@{}>{\hspre}l<{\hspost}@{}}%
\column{16}{@{}>{\hspre}l<{\hspost}@{}}%
\column{E}{@{}>{\hspre}l<{\hspost}@{}}%
\>[B]{}\Varid{...}{}\<[E]%
\\
\>[B]{}\Varid{⟦}\;\Varid{χ}\;{}\<[11]%
\>[11]{}\Varid{⟧}\;\mathrel{=}\;{}\<[16]%
\>[16]{}\Conid{Syn}\;\Varid{χ}\;\Varid{+}\;Ξ_T\;\Varid{χ}{}\<[E]%
\ColumnHook
\end{hscode}\resethooks
\begin{hscode}\SaveRestoreHook
\column{B}{@{}>{\hspre}l<{\hspost}@{}}%
\column{15}{@{}>{\hspre}l<{\hspost}@{}}%
\column{E}{@{}>{\hspre}l<{\hspost}@{}}%
\>[B]{}\Varid{...}{}\<[E]%
\\
\>[B]{}\Varid{⟦}\;\underline{ξ}\;\Varid{⟧}\;Σ_V\;Γ_V\;{}\<[15]%
\>[15]{}\mathrel{=}\;\Varid{⦇}\;(\Varid{inr}\;\Varid{ξ})\;\Varid{⦈}{}\<[E]%
\ColumnHook
\end{hscode}\resethooks

This rather simple change has a practically significant impact:
primitive operations defined in \ensuremath{Σ_V}, can now pattern match on their
input of base type, and provide optimised versions based on the
available values. This is demonstrated in the example presented in
Section \ref{sec:Example}. For clarity, the following datatype can be
used instead of plain sums:
\begin{hscode}\SaveRestoreHook
\column{B}{@{}>{\hspre}l<{\hspost}@{}}%
\column{23}{@{}>{\hspre}l<{\hspost}@{}}%
\column{26}{@{}>{\hspre}l<{\hspost}@{}}%
\column{E}{@{}>{\hspre}l<{\hspost}@{}}%
\>[B]{}\Keyword{data}\;\Conid{PossibleValue}\;\Varid{χ}\;{}\<[23]%
\>[23]{}\mathrel{=}\;{}\<[26]%
\>[26]{}\Conid{Exp}\;(\Conid{Syn}\;\Varid{χ})\;{}\<[E]%
\\
\>[23]{}\mid \;{}\<[26]%
\>[26]{}\Conid{Val}\;(Ξ_T\;\Varid{χ}){}\<[E]%
\ColumnHook
\end{hscode}\resethooks

\subsubsection{Reification}
\label{sec:Smart:Reification}
Reification in Section \ref{sec:Sums:Reification}, is updated to
take into account the convertibility rules:

\begin{hscode}\SaveRestoreHook
\column{B}{@{}>{\hspre}l<{\hspost}@{}}%
\column{11}{@{}>{\hspre}l<{\hspost}@{}}%
\column{E}{@{}>{\hspre}l<{\hspost}@{}}%
\>[B]{}\Varid{↓}\;\mathbin{:}\;\Varid{α}\;\Varid{∼⁺}\;\Conid{A}\;\Varid{→}\;\Varid{α}\;\Varid{→}\;\Conid{Syn}\;\Conid{A}{}\<[E]%
\\
\>[B]{}\Varid{↓}\;\Varid{...}{}\<[E]%
\\
\>[B]{}\Varid{↓}\;(\Varid{⇑⁺ᵣ}\;\Varid{a}\;{}\<[11]%
\>[11]{}\Varid{f})\;\Conid{V}\;\mathrel{=}\;\Varid{↓}\;\Varid{a}\;(\Varid{f}\;\Conid{V}){}\<[E]%
\\[\blanklineskip]%
\>[B]{}\Varid{↑}\;\mathbin{:}\;\Varid{α}\;\Varid{∼⁻}\;\Conid{A}\;\Varid{→}\;\Conid{Syn}\;\Conid{A}\;\Varid{↝}\;\Varid{α}{}\<[E]%
\\
\>[B]{}\Varid{↑}\;\Varid{...}{}\<[E]%
\\
\>[B]{}\Varid{↑}\;(\Varid{⇑⁻ᵣ}\;\Varid{a}\;\Varid{f})\;\Conid{M}\;\mathrel{=}\;\Varid{⦇}\;\Varid{f}\;(\Varid{↑}\;\Varid{a}\;\Conid{M})\;\Varid{⦈}{}\<[E]%
\ColumnHook
\end{hscode}\resethooks

\subsection{Richer Languages}
\label{sec:Richer}
There is a wealth of solutions available in NBE and related areas that
can be adopted to support embedding by normalisation of languages with
features not covered in this paper.

In the context of partial evaluation, Danvy, his collaborators, and others
worked on a series of extensions to the algorithm presented
above. They considered support for object language features such as
recursion, side-effects, syntactic sharing, laziness and memoization, and
datatypes (e.g., see \citet{Pragmatics,Memoization,Online,Sheard}).

In the context of type theory, Altenkirch, Dybjer, and others extended
NBE to richer type systems. They considered systems such as variants
of Martin-Löf type theory (e.g., see \citet{NBE-TT}), polymorphic
lambda calculus \citep{NBE-F}, and simply typed lambda calculus with
strong sums \citep{NBE-Sum,Extensional}.

\subsection{An Example}
\label{sec:Example}
As an example, consider the "hello world" of program generation, the
power function:

\begin{hscode}\SaveRestoreHook
\column{B}{@{}>{\hspre}l<{\hspost}@{}}%
\column{3}{@{}>{\hspre}l<{\hspost}@{}}%
\column{5}{@{}>{\hspre}l<{\hspost}@{}}%
\column{8}{@{}>{\hspre}l<{\hspost}@{}}%
\column{13}{@{}>{\hspre}l<{\hspost}@{}}%
\column{18}{@{}>{\hspre}l<{\hspost}@{}}%
\column{E}{@{}>{\hspre}l<{\hspost}@{}}%
\>[B]{}\Varid{power}\;\mathbin{:}\;\Conid{ℤ}\;\Varid{→}\;\Conid{Syn}\;(\underline{ℚ}\;\underline{→}\;\underline{ℚ}){}\<[E]%
\\
\>[B]{}\Varid{power}\;\Varid{n}\;\mathrel{=}\;\underline{λ}\;\Varid{x}\;\underline{→}\;{}\<[E]%
\\
\>[B]{}\hsindent{3}{}\<[3]%
\>[3]{}\Varid{if}\;\Varid{n}\;\Varid{<}\;\Varid{0}\;{}\<[18]%
\>[18]{}\Varid{then}\;{}\<[E]%
\\
\>[3]{}\hsindent{2}{}\<[5]%
\>[5]{}\underline{\text{if}}\;\Varid{x}\;\underline{==}\;\underline{0}\;{}\<[E]%
\\
\>[3]{}\hsindent{2}{}\<[5]%
\>[5]{}\underline{\text{then}}\;\underline{0}\;{}\<[E]%
\\
\>[3]{}\hsindent{2}{}\<[5]%
\>[5]{}\underline{\text{else}}\;(\underline{-1}\;\underline{/}\;(\Varid{power}\;(\Varid{-}\;\Varid{n})\;\underline{@}\;\Varid{x}))\;{}\<[E]%
\\
\>[B]{}\hsindent{3}{}\<[3]%
\>[3]{}\Varid{else}\;\Varid{if}\;\Varid{n}\;==\;\Varid{0}\;\Varid{then}\;{}\<[E]%
\\
\>[3]{}\hsindent{2}{}\<[5]%
\>[5]{}\underline{1}\;{}\<[E]%
\\
\>[B]{}\hsindent{3}{}\<[3]%
\>[3]{}\Varid{else}\;\Varid{if}\;\Varid{even}\;\Varid{n}\;\Varid{then}\;{}\<[E]%
\\
\>[3]{}\hsindent{2}{}\<[5]%
\>[5]{}({}\<[8]%
\>[8]{}\Keyword{let}\;{}\<[13]%
\>[13]{}\Varid{y}\;\mathrel{=}\;\Varid{power}\;(\Varid{n}\;\mathbin{/}\;\Varid{2})\;\underline{@}\;\Varid{x}{}\<[E]%
\\
\>[8]{}\Keyword{in}\;{}\<[13]%
\>[13]{}\Varid{y}\;\underline{*}\;\Varid{y})\;{}\<[E]%
\\
\>[B]{}\hsindent{3}{}\<[3]%
\>[3]{}\Varid{else}\;{}\<[E]%
\\
\>[3]{}\hsindent{2}{}\<[5]%
\>[5]{}\Varid{x}\;\underline{*}\;(\Varid{power}\;(\Varid{n}\;\Varid{-}\;\Varid{1})\;\underline{@}\;\Varid{x}){}\<[E]%
\ColumnHook
\end{hscode}\resethooks

It takes two arguments and raises the second to the power of the first
argument. Following the convention (e.g., see \citet{QDSL}), it is
written in the ``staged" style: \ensuremath{\Varid{power}} is a meta-function in the host
language that provided integer host value \ensuremath{\Varid{n}} produces object terms of
the type \ensuremath{\underline{ℚ}\;\underline{→}\;\underline{ℚ}}. For pedagogical purposes, we avoid techniques
that further optimise this function but obscure its presentation.

Following the parametric model proposed in this section, for defining
the syntax one only needs to provide the following:

\begin{description}
\item [Base Types], which includes type rational numbers
\begin{hscode}\SaveRestoreHook
\column{B}{@{}>{\hspre}l<{\hspost}@{}}%
\column{E}{@{}>{\hspre}l<{\hspost}@{}}%
\>[B]{}\Conid{X}\;\mathrel{=}\;\{\mskip1.5mu \underline{ℚ}\mskip1.5mu\}{}\<[E]%
\ColumnHook
\end{hscode}\resethooks
\item [Literals], which includes literals of rational numbers
\begin{hscode}\SaveRestoreHook
\column{B}{@{}>{\hspre}l<{\hspost}@{}}%
\column{5}{@{}>{\hspre}l<{\hspost}@{}}%
\column{E}{@{}>{\hspre}l<{\hspost}@{}}%
\>[B]{}Ξ_T\;{}\<[5]%
\>[5]{}\mathrel{=}\;\{\mskip1.5mu \underline{ℚ}\;\Varid{↦}\;\Conid{ℚ}\mskip1.5mu\}{}\<[E]%
\ColumnHook
\end{hscode}\resethooks
\item [Primitives], which includes equality, multiplication, and division
                    operations on rational numbers
\begin{hscode}\SaveRestoreHook
\column{B}{@{}>{\hspre}l<{\hspost}@{}}%
\column{5}{@{}>{\hspre}l<{\hspost}@{}}%
\column{8}{@{}>{\hspre}l<{\hspost}@{}}%
\column{10}{@{}>{\hspre}l<{\hspost}@{}}%
\column{12}{@{}>{\hspre}l<{\hspost}@{}}%
\column{15}{@{}>{\hspre}l<{\hspost}@{}}%
\column{19}{@{}>{\hspre}l<{\hspost}@{}}%
\column{24}{@{}>{\hspre}l<{\hspost}@{}}%
\column{33}{@{}>{\hspre}l<{\hspost}@{}}%
\column{47}{@{}>{\hspre}l<{\hspost}@{}}%
\column{E}{@{}>{\hspre}l<{\hspost}@{}}%
\>[B]{}\Conid{Σ}\;\mathrel{=}\;\{\mskip1.5mu {}\<[8]%
\>[8]{}\underline{==}\;\Varid{↦}\;\Varid{2}\;\Varid{,}{}\<[E]%
\\
\>[8]{}\underline{*}\;{}\<[12]%
\>[12]{}\Varid{↦}\;\Varid{2}\;\Varid{,}{}\<[E]%
\\
\>[8]{}\underline{/}\;{}\<[12]%
\>[12]{}\Varid{↦}\;\Varid{2}\mskip1.5mu\}{}\<[E]%
\\[\blanklineskip]%
\>[B]{}Σ_T\;{}\<[5]%
\>[5]{}\mathrel{=}\;\{\mskip1.5mu {}\<[10]%
\>[10]{}\underline{==}\;{}\<[15]%
\>[15]{}\mathbin{:}\;\{\mskip1.5mu \underline{ℚ}\;\Varid{,}\;\underline{ℚ}\mskip1.5mu\}\;\Varid{↦}\;\underline{\text{Bool}}\;\Varid{,}{}\<[E]%
\\
\>[10]{}\underline{*}\;{}\<[15]%
\>[15]{}\mathbin{:}\;\{\mskip1.5mu \underline{ℚ}\;\Varid{,}\;\underline{ℚ}\mskip1.5mu\}\;\Varid{↦}\;\underline{ℚ}\;\Varid{,}{}\<[E]%
\\
\>[10]{}\underline{/}\;{}\<[15]%
\>[15]{}\mathbin{:}\;\{\mskip1.5mu \underline{ℚ}\;\Varid{,}\;\underline{ℚ}\mskip1.5mu\}\;\Varid{↦}\;\underline{ℚ}\mskip1.5mu\}{}\<[E]%
\\[\blanklineskip]%
\>[B]{}Σ_V\;\mathbin{:}\;\Varid{⦇}\;Σ_T\;\Varid{⦈}{}\<[E]%
\\
\>[B]{}Σ_V\;{}\<[5]%
\>[5]{}\mathrel{=}\;\Varid{⦇}\;\{\mskip1.5mu {}\<[E]%
\\
\>[5]{}\hsindent{5}{}\<[10]%
\>[10]{}(\Conid{Val}\;\Conid{V})\;{}\<[19]%
\>[19]{}\Varid{==ᵥ}\;{}\<[24]%
\>[24]{}(\Conid{Val}\;\Conid{W})\;{}\<[33]%
\>[33]{}\mathrel{=}\;\Varid{⦇}\;\Conid{V}\;==\;\Conid{W}\;\Varid{⦈}{}\<[E]%
\\
\>[5]{}\hsindent{5}{}\<[10]%
\>[10]{}(\Conid{Val}\;\Conid{V})\;{}\<[19]%
\>[19]{}\Varid{==ᵥ}\;{}\<[24]%
\>[24]{}(\Conid{Exp}\;\Conid{N})\;{}\<[33]%
\>[33]{}\mathrel{=}\;\Varid{↑}\;\Conid{Boolᵣ}\;(\underline{V}\;\underline{==}\;\Conid{N}){}\<[E]%
\\
\>[5]{}\hsindent{5}{}\<[10]%
\>[10]{}(\Conid{Exp}\;\Conid{M})\;{}\<[19]%
\>[19]{}\Varid{==ᵥ}\;{}\<[24]%
\>[24]{}(\Conid{Val}\;\Conid{W})\;{}\<[33]%
\>[33]{}\mathrel{=}\;\Varid{↑}\;\Conid{Boolᵣ}\;(\Conid{M}\;{}\<[47]%
\>[47]{}\underline{==}\;\underline{W}){}\<[E]%
\\
\>[5]{}\hsindent{5}{}\<[10]%
\>[10]{}(\Conid{Exp}\;\Conid{M})\;{}\<[19]%
\>[19]{}\Varid{==ᵥ}\;{}\<[24]%
\>[24]{}(\Conid{Exp}\;\Conid{N})\;{}\<[33]%
\>[33]{}\mathrel{=}\;\Varid{↑}\;\Conid{Boolᵣ}\;(\Conid{M}\;{}\<[47]%
\>[47]{}\underline{==}\;\Conid{N})\;\Varid{,}{}\<[E]%
\\[\blanklineskip]%
\>[5]{}\hsindent{5}{}\<[10]%
\>[10]{}(\Conid{Val}\;\Conid{V})\;{}\<[19]%
\>[19]{}\Varid{*ᵥ}\;{}\<[24]%
\>[24]{}(\Conid{Val}\;\Conid{W})\;{}\<[33]%
\>[33]{}\mathrel{=}\;\Varid{⦇}\;\Conid{Val}\;(\Conid{V}\;\Varid{*}\;\Conid{W})\;\Varid{⦈}{}\<[E]%
\\
\>[5]{}\hsindent{5}{}\<[10]%
\>[10]{}(\Conid{Val}\;\Varid{1})\;{}\<[19]%
\>[19]{}\Varid{*ᵥ}\;{}\<[24]%
\>[24]{}(\Conid{Exp}\;\Conid{N})\;{}\<[33]%
\>[33]{}\mathrel{=}\;\Varid{⦇}\;\Conid{Exp}\;\Conid{N}\;\Varid{⦈}{}\<[E]%
\\
\>[5]{}\hsindent{5}{}\<[10]%
\>[10]{}(\Conid{Val}\;\Conid{V})\;{}\<[19]%
\>[19]{}\Varid{*ᵥ}\;{}\<[24]%
\>[24]{}(\Conid{Exp}\;\Conid{N})\;{}\<[33]%
\>[33]{}\mathrel{=}\;\Varid{⦇}\;\Conid{Exp}\;(\underline{V}\;\underline{*}\;\Conid{N})\;\Varid{⦈}{}\<[E]%
\\
\>[5]{}\hsindent{5}{}\<[10]%
\>[10]{}(\Conid{Exp}\;\Conid{M})\;{}\<[19]%
\>[19]{}\Varid{*ᵥ}\;{}\<[24]%
\>[24]{}(\Conid{Val}\;\Varid{1})\;{}\<[33]%
\>[33]{}\mathrel{=}\;\Varid{⦇}\;\Conid{Exp}\;\Conid{M}\;\Varid{⦈}{}\<[E]%
\\
\>[5]{}\hsindent{5}{}\<[10]%
\>[10]{}(\Conid{Exp}\;\Conid{M})\;{}\<[19]%
\>[19]{}\Varid{*ᵥ}\;{}\<[24]%
\>[24]{}(\Conid{Val}\;\Conid{W})\;{}\<[33]%
\>[33]{}\mathrel{=}\;\Varid{⦇}\;\Conid{Exp}\;(\Conid{M}\;\underline{*}\;\underline{W})\;\Varid{⦈}{}\<[E]%
\\
\>[5]{}\hsindent{5}{}\<[10]%
\>[10]{}(\Conid{Exp}\;\Conid{M})\;{}\<[19]%
\>[19]{}\Varid{*ᵥ}\;{}\<[24]%
\>[24]{}(\Conid{Exp}\;\Conid{N})\;{}\<[33]%
\>[33]{}\mathrel{=}\;\Varid{⦇}\;\Conid{Exp}\;(\Conid{M}\;\underline{*}\;\Conid{N})\;\Varid{⦈,}{}\<[E]%
\\[\blanklineskip]%
\>[5]{}\hsindent{5}{}\<[10]%
\>[10]{}(\Conid{Val}\;\Conid{V})\;{}\<[19]%
\>[19]{}\Varid{/ᵥ}\;{}\<[24]%
\>[24]{}(\Conid{Val}\;\Conid{W})\;{}\<[33]%
\>[33]{}\mathrel{=}\;\Varid{⦇}\;\Conid{Val}\;(\Conid{V}\;\mathbin{/}\;\Conid{W})\;\Varid{⦈}{}\<[E]%
\\
\>[5]{}\hsindent{5}{}\<[10]%
\>[10]{}(\Conid{Val}\;\Conid{V})\;{}\<[19]%
\>[19]{}\Varid{/ᵥ}\;{}\<[24]%
\>[24]{}(\Conid{Exp}\;\Conid{N})\;{}\<[33]%
\>[33]{}\mathrel{=}\;\Varid{⦇}\;\Conid{Exp}\;(\underline{V}\;\underline{/}\;\Conid{N})\;\Varid{⦈}{}\<[E]%
\\
\>[5]{}\hsindent{5}{}\<[10]%
\>[10]{}(\Conid{Exp}\;\Conid{M})\;{}\<[19]%
\>[19]{}\Varid{/ᵥ}\;{}\<[24]%
\>[24]{}(\Conid{Val}\;\Varid{1})\;{}\<[33]%
\>[33]{}\mathrel{=}\;\Varid{⦇}\;\Conid{Exp}\;\Conid{M}\;\Varid{⦈}{}\<[E]%
\\
\>[5]{}\hsindent{5}{}\<[10]%
\>[10]{}(\Conid{Exp}\;\Conid{M})\;{}\<[19]%
\>[19]{}\Varid{/ᵥ}\;{}\<[24]%
\>[24]{}(\Conid{Val}\;\Conid{W})\;{}\<[33]%
\>[33]{}\mathrel{=}\;\Varid{⦇}\;\Conid{Exp}\;(\Conid{M}\;\underline{/}\;\underline{W})\;\Varid{⦈}{}\<[E]%
\\
\>[5]{}\hsindent{5}{}\<[10]%
\>[10]{}(\Conid{Exp}\;\Conid{M})\;{}\<[19]%
\>[19]{}\Varid{/ᵥ}\;{}\<[24]%
\>[24]{}(\Conid{Exp}\;\Conid{N})\;{}\<[33]%
\>[33]{}\mathrel{=}\;\Varid{⦇}\;\Conid{Exp}\;(\Conid{M}\;\underline{/}\;\Conid{N})\;\Varid{⦈}\mskip1.5mu\}\;\Varid{⦈}{}\<[E]%
\ColumnHook
\end{hscode}\resethooks
\end{description}

Above relies on the definition of Boolean values defined as a sum of unit types:
\begin{hscode}\SaveRestoreHook
\column{B}{@{}>{\hspre}l<{\hspost}@{}}%
\column{6}{@{}>{\hspre}l<{\hspost}@{}}%
\column{9}{@{}>{\hspre}l<{\hspost}@{}}%
\column{E}{@{}>{\hspre}l<{\hspost}@{}}%
\>[B]{}\underline{\text{Bool}}\;{}\<[9]%
\>[9]{}\mathrel{=}\;\underline{⟨⟩}\;\underline{+}\;\underline{⟨⟩}{}\<[E]%
\\
\>[B]{}\underline{\text{false}}\;{}\<[9]%
\>[9]{}\mathrel{=}\;\underline{\Varid{inl}}\;\underline{⟨⟩}{}\<[E]%
\\
\>[B]{}\underline{\text{true}}\;{}\<[9]%
\>[9]{}\mathrel{=}\;\underline{\Varid{inr}}\;\underline{⟨⟩}{}\<[E]%
\\
\>[B]{}\underline{\text{if}}\;{}\<[6]%
\>[6]{}\Conid{L}\;\underline{\text{then}}\;\Conid{M}\;\underline{\text{else}}\;\Conid{N}\;\mathrel{=}\;\underline{\Varid{case}}\;\Conid{L}\;(\underline{λ}\;\Varid{x}\;\underline{→}\;\Conid{N})\;(\underline{λ}\;\Varid{y}\;\underline{→}\;\Conid{M}){}\<[E]%
\\[\blanklineskip]%
\>[B]{}\Conid{Boolᵣ}\;{}\<[9]%
\>[9]{}\mathrel{=}\;\Varid{⟨⟩ᵣ}\;\Varid{+ᵣ}\;\Varid{⟨⟩ᵣ}{}\<[E]%
\ColumnHook
\end{hscode}\resethooks

Running \ensuremath{\Varid{norm}\;(\Varid{power}\;\Varid{-6})} results in the following code:
\begin{hscode}\SaveRestoreHook
\column{B}{@{}>{\hspre}l<{\hspost}@{}}%
\column{12}{@{}>{\hspre}l<{\hspost}@{}}%
\column{29}{@{}>{\hspre}l<{\hspost}@{}}%
\column{E}{@{}>{\hspre}l<{\hspost}@{}}%
\>[B]{}(\underline{λ}\;\Varid{x₀}\;\underline{→}\;{}\<[12]%
\>[12]{}\underline{\text{if}}\;(\Varid{x₀}\;\underline{==}\;\underline{0})\;{}\<[E]%
\\
\>[12]{}\underline{\text{then}}\;\underline{0}\;{}\<[E]%
\\
\>[12]{}\underline{\text{else}}\;(\underline{-1}\;\underline{/}\;({}\<[29]%
\>[29]{}(\Varid{x₀}\;\underline{*}\;(\Varid{x₀}\;\underline{*}\;\Varid{x₀}))\;\underline{*}{}\<[E]%
\\
\>[29]{}(\Varid{x₀}\;\underline{*}\;(\Varid{x₀}\;\underline{*}\;\Varid{x₀}))))){}\<[E]%
\ColumnHook
\end{hscode}\resethooks

Notice that primitives in \ensuremath{Σ_V} are smart. They are defined by pattern
matching on the inputs, and produce optimised terms based on the
available inputs. For instance, multiplication of a syntactic term \ensuremath{\Conid{M}}
by one, simplifies to \ensuremath{\Conid{M}}. Without such smart primitives, running
\ensuremath{\Varid{norm}\;(\Varid{power}\;\Varid{-6})} results in the following code:

\begin{hscode}\SaveRestoreHook
\column{B}{@{}>{\hspre}l<{\hspost}@{}}%
\column{12}{@{}>{\hspre}l<{\hspost}@{}}%
\column{29}{@{}>{\hspre}l<{\hspost}@{}}%
\column{E}{@{}>{\hspre}l<{\hspost}@{}}%
\>[B]{}(\underline{λ}\;\Varid{x₀}\;\underline{→}\;{}\<[12]%
\>[12]{}\underline{\text{if}}\;(\Varid{x₀}\;\underline{==}\;\underline{0})\;{}\<[E]%
\\
\>[12]{}\underline{\text{then}}\;\underline{0}\;{}\<[E]%
\\
\>[12]{}\underline{\text{else}}\;(\underline{-1}\;\underline{/}\;({}\<[29]%
\>[29]{}(\Varid{x₀}\;\underline{*}\;((\Varid{x₀}\;\underline{*}\;\underline{1})\;\underline{*}\;(\Varid{x₀}\;\underline{*}\;\underline{1})))\;\underline{*}{}\<[E]%
\\
\>[29]{}(\Varid{x₀}\;\underline{*}\;((\Varid{x₀}\;\underline{*}\;\underline{1})\;\underline{*}\;(\Varid{x₀}\;\underline{*}\;\underline{1}))))){}\<[E]%
\ColumnHook
\end{hscode}\resethooks

To demonstrate normalisation of sum types, a simple form of
abstraction is considered: handling corner-cases. The definition of
\ensuremath{\Conid{Power}} is split into two parts. One alters definition of \ensuremath{\Conid{Power}} to
return \ensuremath{\Varid{nothing}} of \ensuremath{\Conid{Maybe}} type  instead of \ensuremath{\Varid{0}} when division by zero
happens, and another replaces \ensuremath{\Varid{nothing}} by \ensuremath{\Varid{0}}:

\begin{hscode}\SaveRestoreHook
\column{B}{@{}>{\hspre}l<{\hspost}@{}}%
\column{3}{@{}>{\hspre}l<{\hspost}@{}}%
\column{5}{@{}>{\hspre}l<{\hspost}@{}}%
\column{8}{@{}>{\hspre}l<{\hspost}@{}}%
\column{12}{@{}>{\hspre}l<{\hspost}@{}}%
\column{15}{@{}>{\hspre}l<{\hspost}@{}}%
\column{19}{@{}>{\hspre}l<{\hspost}@{}}%
\column{E}{@{}>{\hspre}l<{\hspost}@{}}%
\>[B]{}\Varid{power'}\;\mathbin{:}\;\Conid{ℤ}\;\Varid{→}\;\Conid{Syn}\;(\underline{ℚ}\;\underline{→}\;\underline{\text{Maybeℚ}}){}\<[E]%
\\
\>[B]{}\Varid{power'}\;\Varid{n}\;\mathrel{=}\;\underline{λ}\;\Varid{x}\;\underline{→}\;{}\<[E]%
\\
\>[B]{}\hsindent{3}{}\<[3]%
\>[3]{}\Varid{if}\;\Varid{n}\;\Varid{<}\;\Varid{0}\;{}\<[19]%
\>[19]{}\Varid{then}\;{}\<[E]%
\\
\>[3]{}\hsindent{2}{}\<[5]%
\>[5]{}\underline{\text{if}}\;\Varid{x}\;\underline{==}\;\underline{0}\;{}\<[E]%
\\
\>[3]{}\hsindent{2}{}\<[5]%
\>[5]{}\underline{\text{then}}\;{}\<[12]%
\>[12]{}\underline{\text{nothing}}\;{}\<[E]%
\\
\>[3]{}\hsindent{2}{}\<[5]%
\>[5]{}\underline{\text{else}}\;{}\<[12]%
\>[12]{}({}\<[15]%
\>[15]{}(\underline{λ}\;\Varid{y}\;\underline{→}\;\underline{-1}\;\underline{/}\;\Varid{y}){}\<[E]%
\\
\>[15]{}\underline{⟨\$⟩}\;(\Varid{power'}\;(\Varid{-}\;\Varid{n})\;\underline{@}\;\Varid{x}))\;{}\<[E]%
\\
\>[B]{}\hsindent{3}{}\<[3]%
\>[3]{}\Varid{else}\;\Varid{if}\;\Varid{n}\;==\;\Varid{0}\;{}\<[19]%
\>[19]{}\Varid{then}\;{}\<[E]%
\\
\>[3]{}\hsindent{2}{}\<[5]%
\>[5]{}\underline{\text{just}}\;\underline{1}\;{}\<[E]%
\\
\>[B]{}\hsindent{3}{}\<[3]%
\>[3]{}\Varid{else}\;\Varid{if}\;\Varid{even}\;\Varid{n}\;{}\<[19]%
\>[19]{}\Varid{then}\;{}\<[E]%
\\
\>[3]{}\hsindent{2}{}\<[5]%
\>[5]{}({}\<[8]%
\>[8]{}(\underline{λ}\;\Varid{y}\;\underline{→}\;\Varid{y}\;\underline{*}\;\Varid{y}){}\<[E]%
\\
\>[8]{}\underline{⟨\$⟩}\;(\Varid{power'}\;(\Varid{n}\;\mathbin{/}\;\Varid{2})\;\underline{@}\;\Varid{x}))\;{}\<[E]%
\\
\>[B]{}\hsindent{3}{}\<[3]%
\>[3]{}\Varid{else}\;{}\<[E]%
\\
\>[3]{}\hsindent{2}{}\<[5]%
\>[5]{}({}\<[8]%
\>[8]{}(\underline{λ}\;\Varid{y}\;\underline{→}\;\Varid{x}\;\underline{*}\;\Varid{y}){}\<[E]%
\\
\>[8]{}\underline{⟨\$⟩}\;(\Varid{power'}\;(\Varid{n}\;\Varid{-}\;\Varid{1})\;\underline{@}\;\Varid{x})){}\<[E]%
\\[\blanklineskip]%
\>[B]{}\Varid{power''}\;\mathbin{:}\;\Conid{ℤ}\;\Varid{→}\;\Conid{Syn}\;(\underline{ℚ}\;\underline{→}\;\underline{ℚ}){}\<[E]%
\\
\>[B]{}\Varid{power''}\;\Varid{n}\;\mathrel{=}\;\underline{λ}\;\Varid{x}\;\underline{→}\;\underline{\text{maybe}}\;(\underline{λ}\;\Varid{z}\;\underline{→}\;\Varid{z})\;\underline{0}\;(\Varid{power'}\;\Varid{n}\;\underline{@}\;\Varid{x}){}\<[E]%
\ColumnHook
\end{hscode}\resethooks

Above relies on the definition of \ensuremath{\Conid{Maybe}} values of rational numbers
defined as a sum type:
\begin{hscode}\SaveRestoreHook
\column{B}{@{}>{\hspre}l<{\hspost}@{}}%
\column{15}{@{}>{\hspre}l<{\hspost}@{}}%
\column{E}{@{}>{\hspre}l<{\hspost}@{}}%
\>[B]{}\underline{\text{Maybeℚ}}\;{}\<[15]%
\>[15]{}\mathrel{=}\;\underline{ℚ}\;\underline{+}\;\underline{⟨⟩}{}\<[E]%
\\
\>[B]{}\underline{\text{just}}\;\Varid{x}\;{}\<[15]%
\>[15]{}\mathrel{=}\;\underline{\Varid{inl}}\;\Varid{x}{}\<[E]%
\\
\>[B]{}\underline{\text{nothing}}\;{}\<[15]%
\>[15]{}\mathrel{=}\;\underline{\Varid{inr}}\;\underline{⟨⟩}{}\<[E]%
\\
\>[B]{}\underline{\text{maybe}}\;\Conid{M}\;\Conid{N}\;\Conid{L}\;{}\<[15]%
\>[15]{}\mathrel{=}\;\underline{\Varid{case}}\;\Conid{L}\;(\underline{λ}\;\Varid{x}\;\underline{→}\;\Conid{M}\;\underline{@}\;\Varid{x})\;(\underline{λ}\;\Varid{y}\;\underline{→}\;\Conid{N}){}\<[E]%
\\
\>[B]{}\Conid{L}\;\underline{⟨\$⟩}\;\Conid{M}\;{}\<[15]%
\>[15]{}\mathrel{=}\;\underline{\text{maybe}}\;(\underline{λ}\;\Varid{x}\;\underline{→}\;\underline{\text{just}}\;(\Conid{L}\;\underline{@}\;\Varid{x}))\;\underline{\text{nothing}}\;\Conid{M}{}\<[E]%
\ColumnHook
\end{hscode}\resethooks

Running \ensuremath{\Varid{norm}\;(\Varid{power''}\;\Varid{-6})} results in exactly the same value as \ensuremath{\Varid{norm}\;(\Varid{power}\;\Varid{-6})}. Normalisation removes the unnecessary code in \ensuremath{\Varid{power''}},
and makes it behave as if no additional layer of abstraction has been
introduced in the first place; as demonstrated, normalisation in EBN
achieves abstraction-without-guilt, even for sum types.

Besides the implementation in Agda, examples and the corresponding EBN
technique in this subsection is implemented in Haskell and is
available at
\url{https://github.com/shayan-najd/Embedding-By-Normalisation/blob/master/Power.hs}.

\section{Discussion \& Related Work}
\label{sec:RelatedWork}
As with any other paper describing correspondence between two areas,
this paper introduces the main body of the related works, while
explaining the related concepts. Focus of this section is to mention
key related areas, besides NBE, and where EBN stands in comparison.

\subsection{Normalised EDSLs}
The work by \citet{svenningsson:combiningJournal} is perhaps the most
closely related to what is presented in this paper. They provide a way
to embed languages which combines deep and shallow embeddings which
allows DSLs to be normalised by using evaluation in the host language.
Phrased in the framework of Embedding by Normalisation, their
methodology matches the form of embedding presented in section
\ref{sec:Basic}. They limit their system to a first-order fragment, to
produce efficient and computationally predictable C code.  They can
use host language functions and products for their DSL
implementations. Though they cannot deal with arbitrary sum types,
although they provide tricks for dealing with a restricted form of sum
type, such as the \ensuremath{\Conid{Maybe}} type in the Haskell standard library. Their
implementation in Haskell uses a type-class which contains two methods
for converting from shallow embedding to deep embedding of terms and
vice versa. The type-class and its instances correspond to the
reification function in EBN, where conversion from shallow to deep can
be seen as reification function and conversion from deep to shallow as
reflection function.
Examples of DSLs which use this style of embedding are Feldspar
\citep{FELDSPAR}, Obsidian \citep{svensson2011obsidian}
and Nikola \citep{Mainland:2010}.

Other important related works, are series of successful EDSLs
implemented in Scala using LMS \citep{scalalms,rompf2012lightweight}.
They use evaluation mechanism of the host language for optimising
DSLs.  Their systems are based on a form staged computation (e.g., see
\citet{metaml}), and they do so by a smart type-directed approach
utilising virtualisation (e.g., see \citet{rompf2013scala}). Rompf has
characterised the essence of LMS, as an approach based on the
two-level lambda calculi (e.g., see \citep{Nielson-2005}).  As explored by
\citet{TDPE}, NBE and two-level calculi are related.  One
interesting future work is to exploit the relation to compare EBN with
the technique underlying LMS.

There are also large bodies of works on embedding specific DSLs in
Haskell that use the evaluation mechanism of Haskell to optimise
embedded terms.  \citet{Gill:CACM} provides a general overview of
embedding techniques in Haskell, and a crisp explanation of the
reification problems addressed in this paper. One possible explanation
of why reification for sum types or primitives appeared difficult (if
not impossible), is that DSL designers presumed the semantic domain to be
a simple sub-set of the host language without continuations or lifted
base types, i.e., the one in Section \ref{sec:Basic:Sem}.
As EBN reveals, to be able to reify terms involving sums
or primitives, one needs to settle for an alternative semantic domain.

\subsection{Partial Evaluation}

As mentioned in Section \ref{sec:Richer}, Danvy's Type-Directed
Partial Evaluation \citep{TDPE} and its extensions are central to the
solutions discussed in this paper.  Partial evaluation comes in two
flavours: offline and online.  Section \ref{sec:Sums} basically
describes an offline partial evaluator, and Section \ref{sec:Smart}
describes an online partial evaluator, though in a limited form.
For a more practical use of online partial evaluation in embedding, see
\citet{leissa2015shallow}.
\citet{DybjerF00,Filinski} characterise the relation between partial
evaluation and NBE.

\subsection{Stand-Alone DSLs}
DSLs can also be implemented as a stand-alone language.  In theory,
for a stand-alone DSL one needs to implemented all the required
machinery, and they do not integrate easily with other
languages. However, there are wide range of tools and frameworks
available that automate parts of the implementation process (e.g., see
\citet{spoofax}). While obviously EBN does not apply to stand-alone
language, the original NBE techniques can definitely be used as a way
to write normalisers for stand-alone DSLs. In theory, it is even
possible to implement tools to automate part of the process.

\section{Conclusion}
\label{sec:conclusion}
Girard, starts the first chapter of his popular book ``Proofs and
Types'' \citep{Girard} by the following paragraph:
\begin{displayquote}
Theoretical Computing is not yet a science. Many basic concepts have not been
clarified, and current work in the area obeys a kind of “wedding cake” paradigm:
for instance language design is reminiscent of Ptolomeic astronomy — forever
in need of further corrections. There are, however, some limited topics such as
complexity theory and denotational semantics which are relatively free from this
criticism.
\end{displayquote}

This paper shows that theoretical Normalisation-By-Evaluation (NBE)
techniques, commonly used in denotational semantics, correspond to
popular embedding techniques, commonly used in programming practice.

This paper characterises the correspondence, and puts it into practice,
by an approach dubbed as Embedding-By-Normalisation (EBN). Then, the
paper employs EBN to clarify some of the basic concepts in the
practical embedding techniques, concepts such as code extraction
(reification) and normalisation. The clarification offered by EBN helps
to solve the problem of extracting object code from embedded programs
involving sum types, such as conditional expressions, and primitives,
such as literals and operations on them.

One final observation of this paper might be that there is science and
beauty at the core of the embedding techniques, but it demands rigour
and patience to uncover.

\paragraph*{Acknowledgements}
Najd was funded by a Google Europe Fellowship in Programming
Technology. Svenningsson was funded by the Swedish Foundation for
Strategic Research under grant RawFP. Lindley and Wadler were funded
by EPSRC Grant EP/K034413/1.

\bibliographystyle{abbrvnamed}
\bibliography{paper}

\end{document}